\documentclass[a4paper,fleqn]{cas-dc}

\usepackage[numbers]{natbib}
\usepackage{graphicx}
\usepackage{subcaption}
\usepackage{commath}
\usepackage[none]{hyphenat}
\usepackage{pbalance}
\usepackage{ulem}

\begin{document}
\let\WriteBookmarks\relax
\def\floatpagepagefraction{1}
\def\textpagefraction{.001}
\let\printorcid\relax

\shorttitle{}
\shortauthors{Haozhe Sun, Zhe Wang and Shaomin Chen}

\title [mode = title]{Potassium-40 geoneutrino detection and the Earth’s large-scale structures imaging by directional geoneutrino detection}                      

\author[1,2,3]{Haozhe Sun}[]
\author[1,2,3]{Zhe Wang}[]
\cormark[1]
\cortext[1]{Corresponding author}
\ead{wangzhe-hep@mail.tsinghua.edu.cn}
\author[1,2,3]{Shaomin Chen}[]

\credit{Conceptualization of this study, Methodology, Software}

\affiliation[1]{organization={Department~of~Engineering~Physics, Tsinghua~University},
                city={Beijing},
                postcode={100084}, 
                country={China}}
\affiliation[2]{organization={Center for High Energy Physics, Tsinghua~University},
                city={Beijing},
                postcode={100084}, 
                country={China}}
\affiliation[3]{organization={Key Laboratory of Particle \& Radiation Imaging (Tsinghua University), Ministry of Education},
                city={Beijing},
                postcode={100084}, 
                country={China}}

\begin{abstract}
Geoneutrinos, the electron (anti)neutrinos generated in decays or decay chains of radioactive elements within the Earth, primarily $\rm ^{40}K$, $\rm ^{238}U$, and $\rm ^{232}Th$, serve as a unique probe of the Earth's inner chemical composition.
A directional geoneutrino detection method with a Cherenkov liquid scintillator is investigated in this work.
Neutrino-electron elastic scattering in the Cherenkov scintillator is employed to detect geoneutrinos. 
The direction reconstruction resolution for neutrinos is studied. The radioactive and solar neutrino backgrounds are taken into account.
The intrinsic neutrino background from the Sun is suppressed with an optimized solar angle cut. 
The required exposure to reach a $3~\sigma$ sensitivity to discover geoneutrinos in a potassium-dominated energy region is 6.9~kiloton-years.
The potential to image Earth’s large-scale structures using directional geoneutrino information is also studied.
The required exposure to reject a uniform terrestrial angle distribution of geoneutrinos at the $3~\sigma$ level exceeds 27~kiloton-years.
\end{abstract}

\begin{keywords}
geoneutrino \sep Cherenkov liquid scintillator \sep potassium-40 geoneutrino \sep Earth imaging
\end{keywords}

\maketitle

\section{Introduction}

Geoneutrinos originate from the decay of radioactive elements within the Earth. They primarily come from three radioactive elements: uranium, thorium, and potassium (U, Th, and K).
Measuring the fluxes of various geoneutrinos and imaging the Earth’s large-scale structures will provide us with a wealth of knowledge~\cite{Bill_2013}.
First, the geoneutrino flux corresponds directly with the radiogenic heat flux. Subtracting it from the total terrestrial surface heat flux (about 46 TW), we can infer the residual heat from the Earth’s primordial gravitational collapse~\cite{Bill_2020}. 
Second, U and Th are refractory elements that condense out of a nebular disk at high temperatures and are empirically observed in equal proportions in chondrites. K is a moderately volatile element that follows a distinct condensation pathway. By measuring the potassium-40 geoneutrinos and knowing the abundance of K in the Earth, we can establish a volatilization curve for this planet, which allows us to infer the abundances of other volatile elements. Comparing these results with those from chondrite meteorites may provide more information in reconstructing the Earth’s accretion and evolution history~\cite{Bill_2020}. 
Third, if the geoneutrinos can be used to image the planet’s large-scale structure, we can gain deeper insights into the Earth. For instance, regarding the Qinghai-Tibet Plateau, the subduction of the Indian tectonic plate has formed a thick crust, and the affinity of U, Th, and K with silicate (lithophile)  leads to the enrichment of heat-generating elements in the shallow geological layers. Models predict a very high flux of geoneutrinos in this region~\cite{Sramek_2016, Wan_2016}. 
With geoneutrino imaging of large-scale structures, we can study these models and reveal whether other unknown structures exist. Furthermore, if we can measure the flux of geoneutrinos from the mantle, we will gain valuable insights into the forces driving plate tectonics and the Earth’s dynamo.

KamLAND experiment reported the first measurement of geoneutrinos in 2005~\cite{KamLAND_2005}. 
The most recent measurements are from KamLAND~\cite{KamLAND_2011}, Borexino~\cite{Borexino_2020}, SNO+~\cite{SNO_2025} and JUNO~\cite{JUNO} experiments. 
The geoneutrinos were detected through the inverse beta decay (IBD) process in large liquid scintillator detectors,
\begin{equation}
        \bar{\nu}_e + p \rightarrow e^+ + n.
    \label{equ:IBD}
\end{equation}
In the IBD reaction, an electron antineutrino interacts with a free proton, producing a positron and a neutron. The positron deposits its kinetic energy and immediately annihilates with an electron, forming the prompt signal. The neutron subsequently thermalizes and is eventually captured by a nucleus, releasing gamma rays, forming a delayed signal.
The temporal and spatial correlation between the prompt and delayed signals provides an effective selection criterion to suppress backgrounds.

The IBD process has a detection threshold of 1.806~MeV, which is higher than the endpoint energy of the $\rm ^{40}K$ geoneutrino spectrum, at 1.31~MeV.
Only uranium and thorium geoneutrinos can be detected.
The directional information obtained from the IBD process is also quite limited~\cite{IBD}.

Previously, it was proposed to use a Cherenkov liquid scintillator\cite{Li_2016, Guo_2017}  to measure the energy and direction of the recoil electrons  from the neutrino-electron elastic scattering (ES) process~\cite{Wang_2020}: 
\begin{equation}
        \nu + e \rightarrow \nu + e.
    \label{equ:ES}
\end{equation}
The ES process can occur with all neutrino flavors and has no interaction threshold.
The recoil electrons retain some directional information of the incoming neutrinos.
These features are quite useful for the discovery of potassium-40 geoneutrinos and for suppressing the intrinsic background of solar neutrinos.
The directionality is also a basic requirement for the geoneutrino Earth imaging.
There have been several new developments in the optional liquid scintillation solvents and fluorescent solutes~\cite{Steven_2020, Steiger}, as well as the reconstruction methods~\cite{Luo_2023}.
In this paper, we revisit the previous idea using the Cherenkov liquid scintillator. 
One problem has been found in the previous simulation work. The candidate selection criteria and sensitivity calculation method have also been upgraded. 
We found an improved sensitivity for detecting potassium-40 geoneutrinos, and it is also possible to generate a large-scale structure image of the Earth.

The paper is organized as follows.
In Sec.~\ref{sec:Signal}, the signal and background processes are described. Sec.~\ref{sec:Simulation} presents the response of the Cherenkov liquid scintillator detector.
Sec.~\ref{sec:Criteria} presents the upgraded selection criteria.
The sensitivities are calculated and presented in Sec.~\ref{sec:Sensitivity}, and the paper is summarized in Sec.~\ref{sec:Summary}.

\section{Geoneutrino signal and backgrounds}
\label{sec:Signal}

This section will explain the geoneutrino signal and background prediction.
The experimental condition is set according to the proposal of the Jinping Neutrino Experiment (JNE)~\cite{JNE}, located at the China Jinping Underground Laboratory (CJPL) in Sichuan Province. 
The CJPL benefits from the 2400-m vertical rock overburden and the 1000-km distance to the nearby commercial reactors.

For the geoneutrino prediction, details of the coordinate system, geoneutrino spectrum, Earth model, neutrino oscillations, and the final kinetic energy spectra of recoil electrons are presented. 

For the CJPL, the cosmic-ray muon flux is about 200 times lower than that of the Borexino experiment, and the related muon-induced background can be vetoed to a negligible level. 
In the energy region of interest, there are reactor neutrinos and solar neutrinos.
The reactor neutrino background at the CJPL is also low, and its flux can be measured well with the IBD process and eventually statistically subtracted.
Solar neutrinos are one of the major backgrounds, which will be presented in detail.
Radioactive backgrounds are also quite important. External radioactive background, i.e., from detector components outside of the liquid scintillator region, can be vetoed by a fiducial volume cut like in the Borexino experiment~\cite{BorexinoPhaseI}.
The contamination of the liquid scintillator itself causes internal radioactive background.
The internal background is irreducible, and a thorough analysis of its impact on the geoneutrino study is carried out.

\subsection{Coordinate system}
\begin{figure}
    \centering
    \includegraphics[width=\columnwidth]{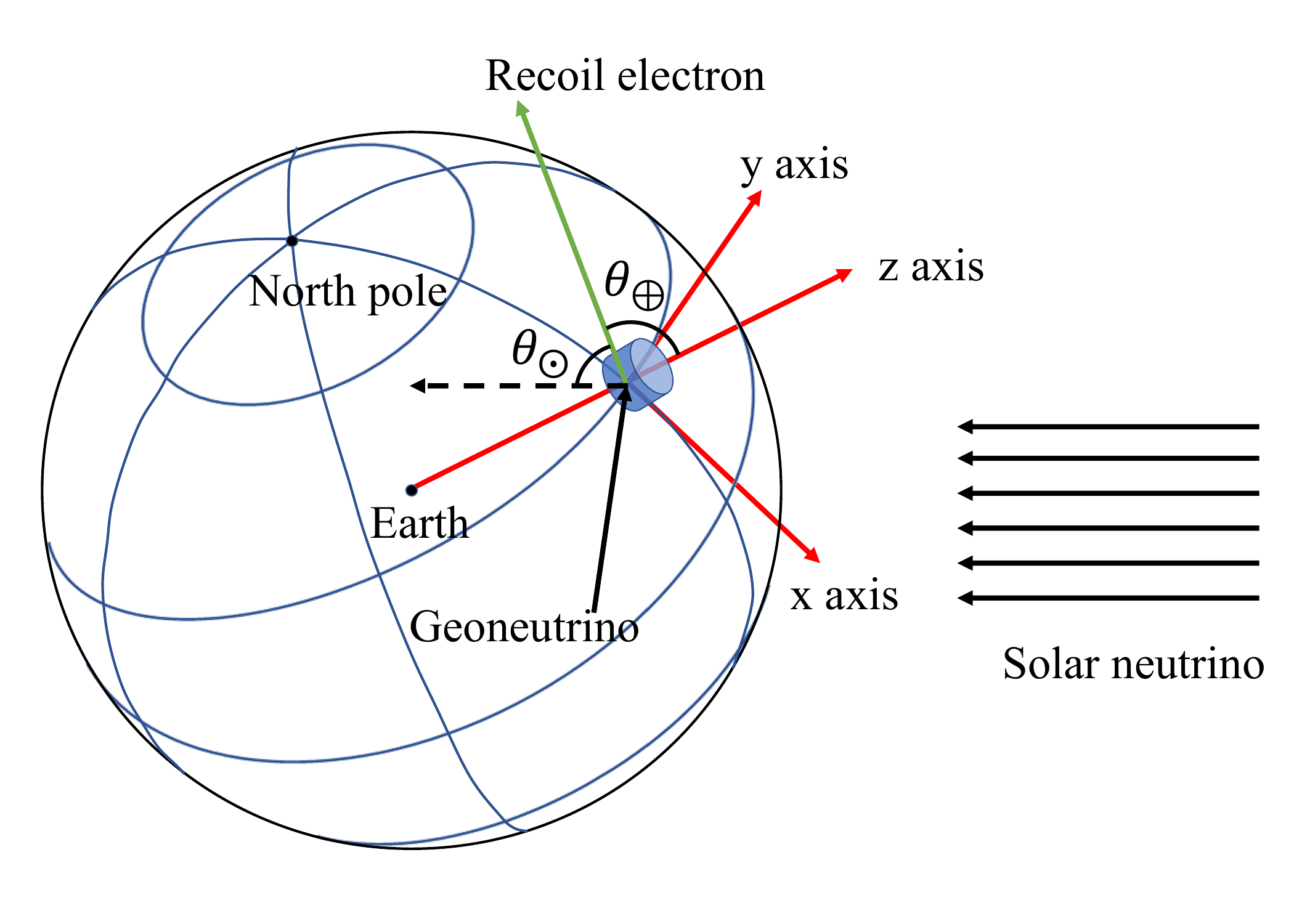}
    \caption{Coordinate systems. The solar coordinate system varies with the relative position of the Earth and the Sun, while the terrestrial coordinate system is fixed on the Earth.}
    \label{fig:CoordinateSystem}
\end{figure}

The directionality is significant in distinguishing the geoneutrino signal from the solar neutrino background. The geoneutrino signal primarily originates from the Earth's interior, whereas the solar neutrino background is strongly correlated with the Sun's direction.
To characterize the difference, two sets of coordinate systems are established. 
The solar coordinate system focuses on the relationship with the solar azimuth, with the solar angle $\theta_\odot$ representing the angle between the recoil electron's direction and the Sun-Earth direction. The terrestrial coordinate system is fixed on the Earth, with the origin set at the detector position, the $z$-axis aligned along the direction from the Earth's center to the detector, and the $x$-axis aligned along the latitudinal direction and pointing towards the south. The angular coordinates $(\theta_\oplus, \phi_{\oplus})$ are defined in this coordinate system. The coordinate systems are illustrated in Fig.~\ref{fig:CoordinateSystem}.

\subsection{Geoneutrino prediction}
The total geoneutrino energy spectrum at the CJPL of flavor $\alpha$ ($\alpha=e$, $\mu$, and $\tau$), originating from an isotope $i$ ($i=$ U, Th, and K), can be written as the following formula.
\begin{equation}
    \begin{aligned}
        \phi_{\alpha i} (E) = &\frac{X_i \lambda_i N_A}{\mu_i} n_{\nu i} S_i(E)  \\
        &\times \int \frac{1}{4\pi L^2}P_{e \alpha}(E, L) \rho(\vec{r}) A_i(\vec{r}) {\rm d^3}\vec{r},    \\
    \end{aligned}
\end{equation}
where $X_i$ represents the mole fraction of an isotope $i$, $\lambda_i$ represents the decay constant for $i$, $N_A$ is the Avogadro's constant, $\mu_i$ is the molar mass of the isotope $i$, $n_{\nu i}$ represents the number of neutrinos emitted in a decay (chain), $S_i(E)$ is the normalized neutrino spectrum of isotope $i$, $L$ is the distance from the location $\vec{r}$ to Jinping detector, $\rho(\vec{r})$ is the density, and $A(\vec{r})$ represents the abundance of the element.
$P_{e \alpha}(E, L)$ is the appearance probability of neutrino flavor $\alpha$ with energy $E$ after an electron neutrino traveling through a distance $L$. 
Neutrinos and antineutrinos of all three flavors contribute to the neutrino-electron elastic scattering process.

\subsection{Geoneutrino spectrum}

\begin{figure}
    \centering
    \includegraphics[width=\columnwidth]{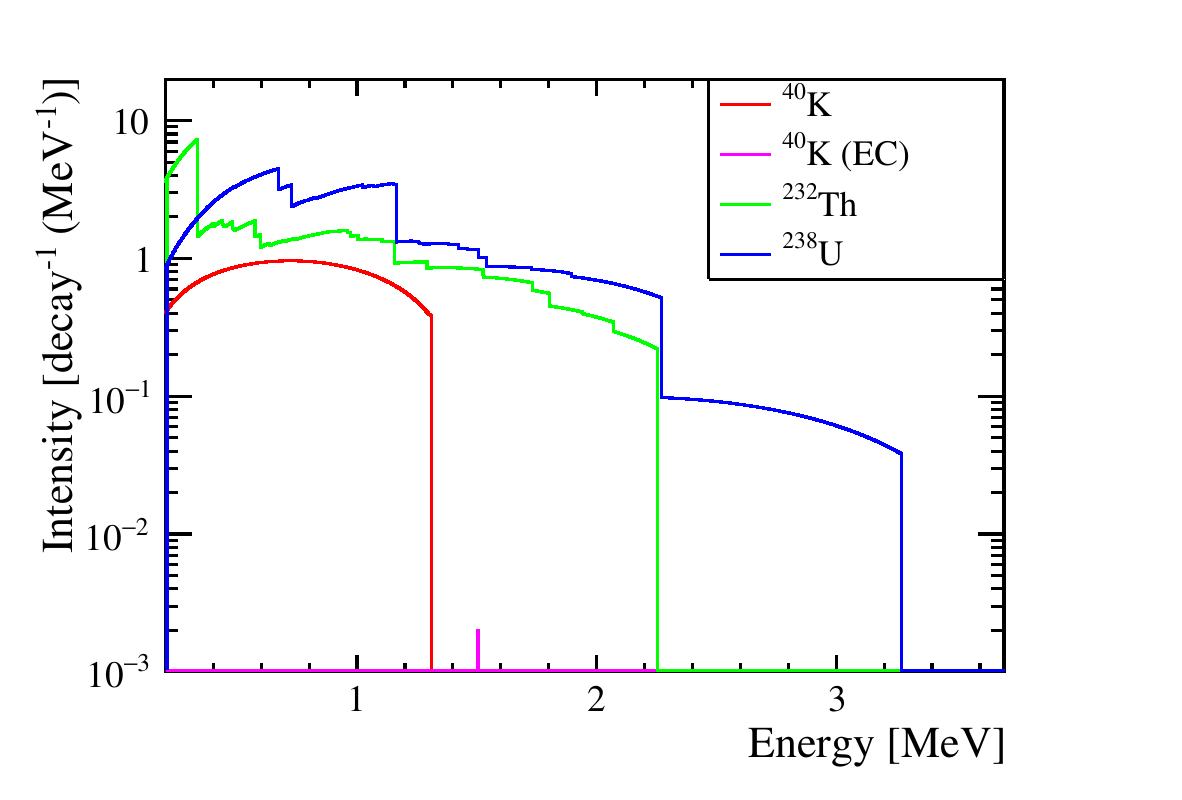}
    \caption{Antineutrino spectrum calculated by Enomoto~\cite{Enomoto_2005}. The monoenergetic peak of $\rm ^{40}K$ orbital electron capture is also shown.}
    \label{fig:NeutrinoSpectrum}
\end{figure}

The geoneutrinos are primarily from uranium-238, thorium-232, and potassium-40:
\begin{equation}
    \begin{aligned}
        {\rm ^{238}U} &\rightarrow {\rm ^{206}Pb} + 8 \alpha + 6 e^- + 6 \bar{\nu}_e + 57.1~{\rm MeV},   \\
        {\rm ^{232}Th} &\rightarrow {\rm ^{208}Pb} + 6 \alpha + 4 e^- + 4 \bar{\nu}_e + 42.7~{\rm MeV},  \\
        {\rm ^{40}K} &\rightarrow {\rm ^{40}Ca} + e^- + \bar{\nu}_e + 1.33~{\rm MeV}~(89.3\%),           \\
        {\rm ^{40}K} + e^- &\rightarrow {\rm ^{40}Ar} + e^- + \nu_e + 1.505~{\rm MeV}~(10.7\%).          \\
    \end{aligned}
    \label{equ:Decay}
\end{equation}
The neutrino spectrum is from the result published by Enomoto in 2005~\cite{Enomoto_2005}. Additionally, the orbital electron capture is considered in this research. Fig.~\ref{fig:NeutrinoSpectrum} shows the (anti)neutrino spectrum from each decay.
Although only 0.2\% of $\rm ^{40}K$ decays via orbital electron capture and emits an electron neutrino with an energy of 1.5~MeV~\cite{BIPM_2010}, these neutrinos contribute approximately 4.5\% of the signal in the energy range dominated by $\rm ^{40}K$ due to the relatively larger cross-section than antineutrinos.

\subsection{Earth model}
\begin{figure}
    \centering
    \begin{subfigure}{\columnwidth}
        \centering
        \includegraphics[width=\columnwidth]{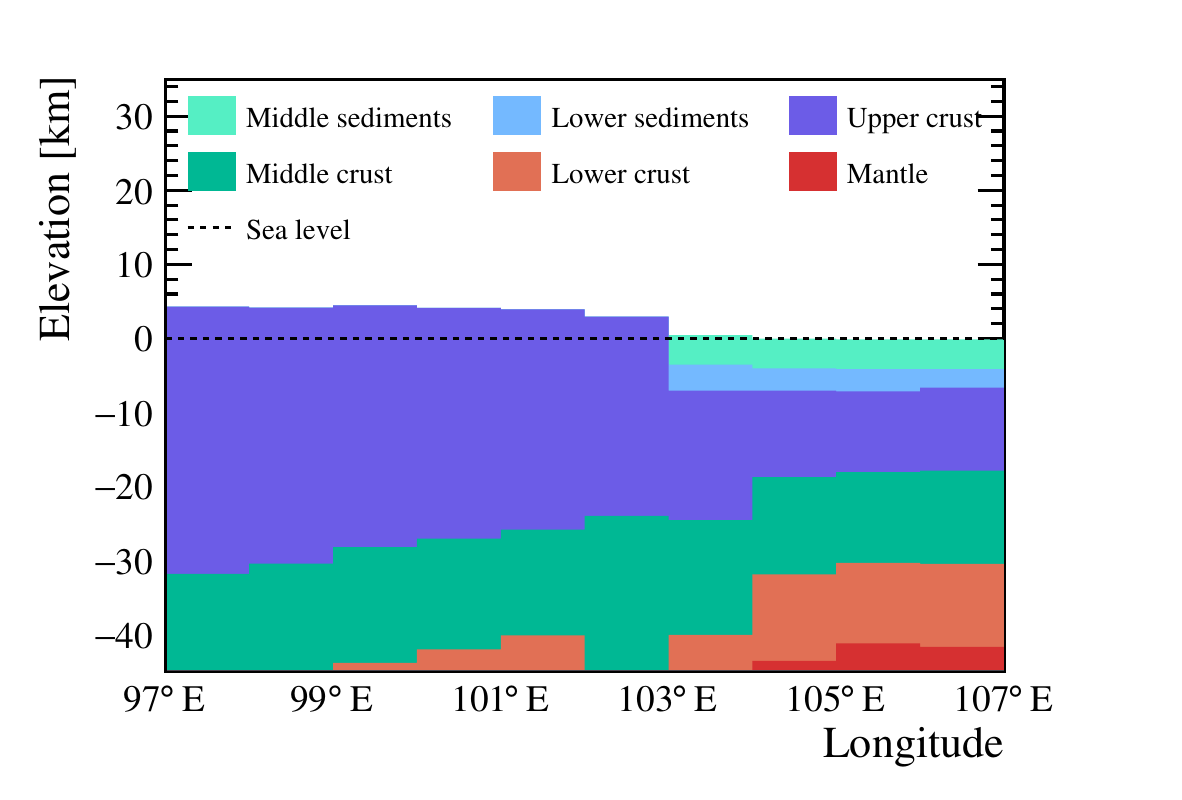}
        \caption{}
        \label{fig:SchematicDigram_North}
    \end{subfigure}
    \begin{subfigure}{\columnwidth}
        \centering
        \includegraphics[width=\columnwidth]{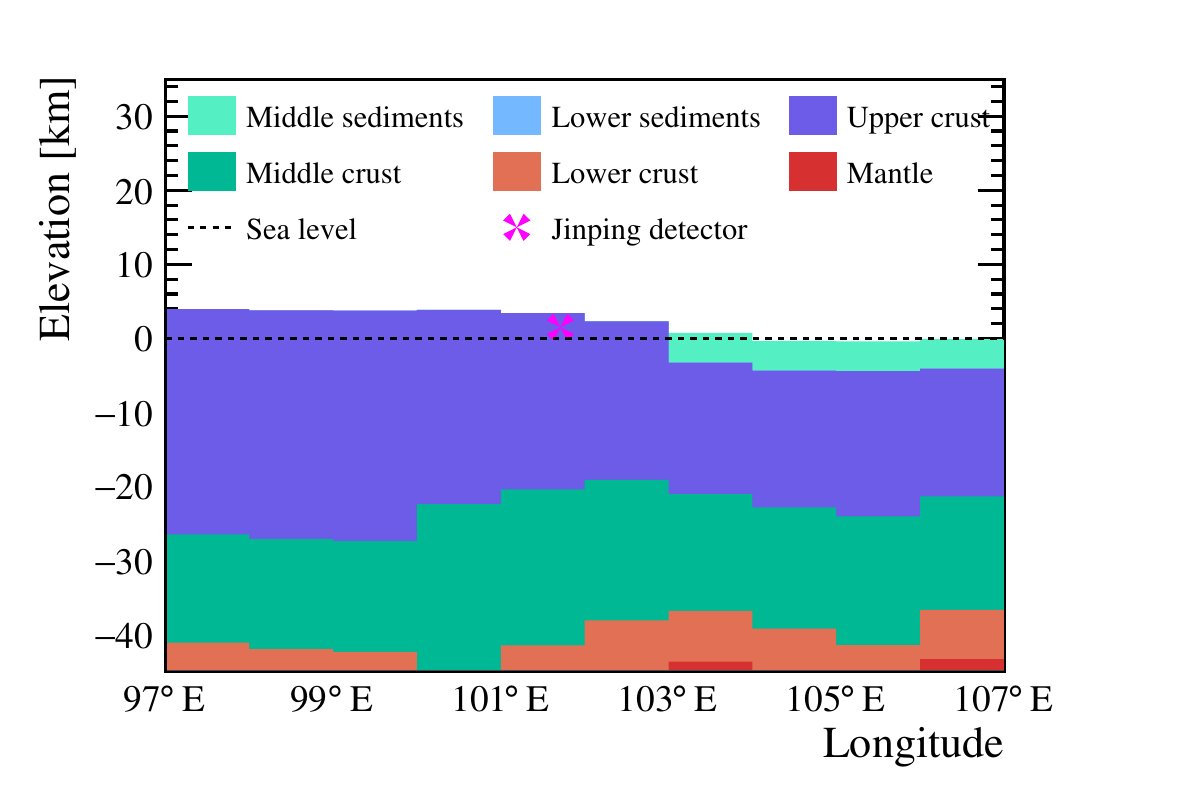}
        \caption{}
        \label{fig:SchematicDigram_Jinping}
    \end{subfigure}
    \begin{subfigure}{\columnwidth}
        \centering
        \includegraphics[width=\columnwidth]{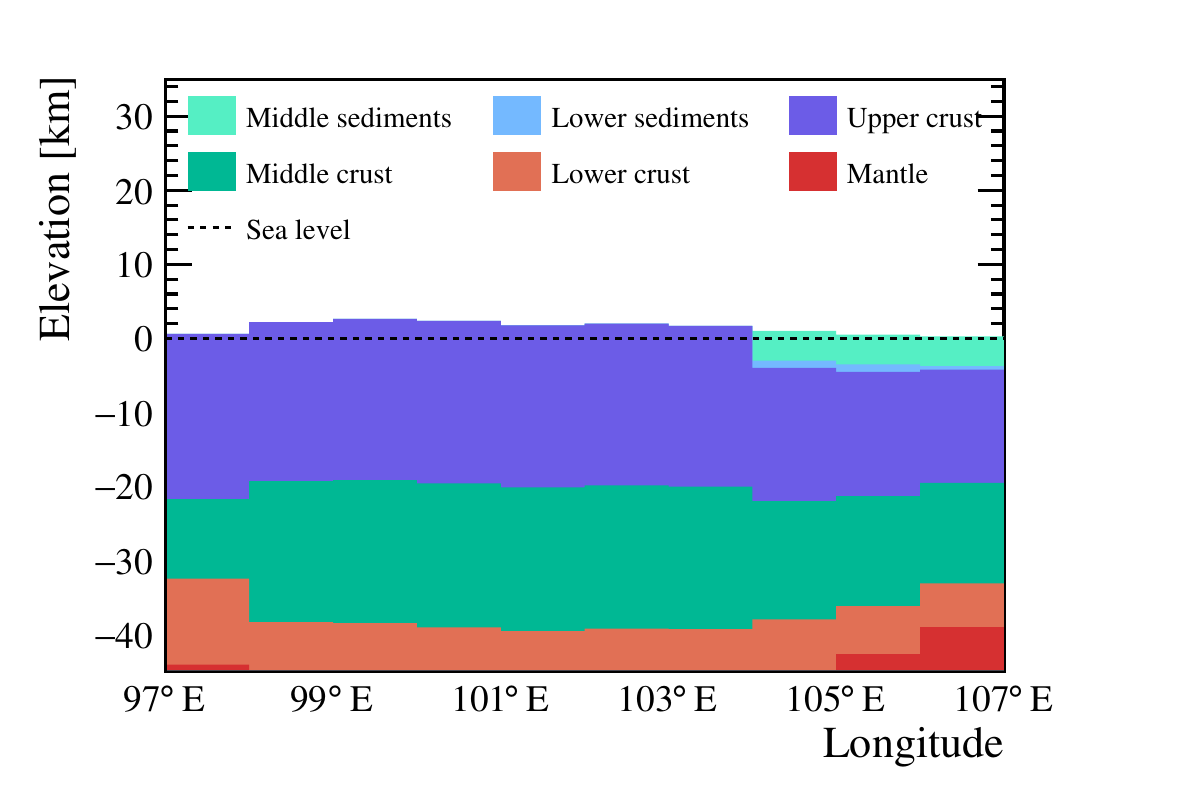}
        \caption{}
        \label{fig:SchematicDigram_South}
    \end{subfigure}
    \caption{
Schematic profiles
along the longitudinal-elevation plane.
Panel (b) shows the profile at the latitude and longitude of CJPL. Panel (a) and (c) plots
are for $2^\circ$ (220 km) north and south of the CJPL, respectively.
}
    \label{fig:SchematicDigram}
\end{figure}

The Crust 1.0 is a global crustal model, which provides a detailed 3D representation of the Earth's crust~\cite{CRUST1.0_2013}. The model is structured as a $1^\circ \times 1^\circ$ grid in latitude and longitude, covering the entire globe. At each grid cell, the crust is divided into 9 geological layers: ice, water, upper, middle, and lower sediments, and upper, middle, and lower crust. For each layer in every grid cell, Crust 1.0 provides key parameters, including boundary depths, density, and crustal type. Because of the relatively high concentration of the radioactive elements in the crust and their proximity to the detector, crust-originated neutrinos account for approximately 70\% of all geoneutrino signals. Beneath the crust lie the continental lithospheric mantle (CLM), depleted mantle (DM), and enriched mantle (EM). 
The density and boundary positions of the CLM, DM, and EM are assumed to take the values from Ref.~\cite{Huang_2013}.
Fig.~\ref{fig:SchematicDigram} shows three schematic profiles along the longitudinal-elevation plane.
Panel (b) shows the profile at the latitude and longitude of the CJPL.
Panels (a) and (c) are for $2^\circ$ north and south of the CJPL, respectively.
The thick crust of the Tibet area is visible.

While the above model defines the geometric and physical structural features of the crust and mantle, the radioactive element abundances are determined based on a medium-Q assumption for the Bulk Silicate Earth~\cite{Sramek_2016}. The abundances of U, Th, and K are assumed to be uniform in a layer.

\subsection{Neutrino oscillation}

Because of neutrino mixing, neutrinos change flavor as they travel in a vacuum or matter. The survival probability of an electron antineutrino with energy $E$ traveling a distance $L$ in a vacuum can be written as:
\begin{equation}
        P_{ee}(E, L) = \sum_{i=1}^3 \left|U_{ei}^\dag U_{ie} {\mathrm e}^{-{\mathrm i} m_i^2 / 2 E} \right|^2,
    \label{equ:SurvivalProbability1}
\end{equation}
where $U$ is the Pontecorvo--Maki--Nakagawa--Sakata (PMNS) matrix~\cite{Gribov_1969,Pontecorvo_1971}.
The survival probability of electron neutrinos is
\begin{equation}
    \begin{aligned}
        P_{ee}(E, L) = 1 &- \cos^4 \theta_{13} \sin^2 2\theta_{12} \sin^2 2\Delta_{21}    \\
        &- \sin^2 2\theta_{13} \cos^2 \theta_{12} \sin^2 \Delta_{31}    \\
        &- \sin^2 2\theta_{13} \sin^2 \theta_{12} \sin^2 \Delta_{32},   \\
    \end{aligned}
    \label{equ:SurvivalProbability2}
\end{equation}
where $\Delta_{ij} = 1.257 \times 10^{-3} \frac{\Delta m_{ij}^2 L}{E}$ and $m_{ij}^2 = |m_j^2 - m_i^2|$, with $\Delta m_{ij}^2$ in $\rm eV^2$, $L$ in km, and $E$ in MeV.
The matter effect is taken into account~\cite{Wolfenstein_1978, Mikheev_1985}.
The oscillation effect is integrated over the whole Earth modeled with the Preliminary Reference Earth Model (PREM)~\cite{PREM_1981}. 

The distance between 1 degree of latitude is about 111~km, which is larger than the typical oscillation length, which is approximately 30~km at 1~MeV for the $\Delta m_{12}^2$ term. For a precise computation of $P_{e\alpha}(E, L)$, the nearby grid cells around the CJPL are divided into more matter-uniform sub-cells. 

\begin{figure}
    \centering
    \includegraphics[width=\columnwidth]{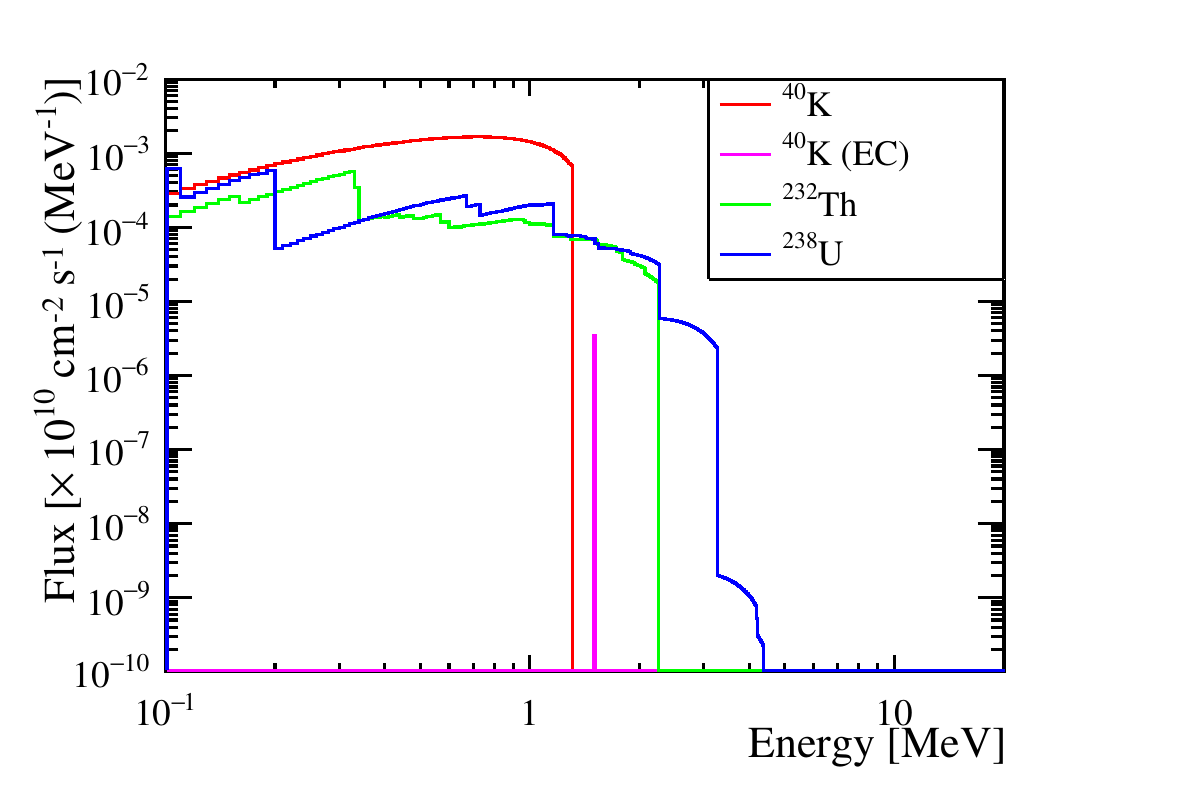}
    \caption{Electron (anti)neutrino flux prediction at Jinping.}
    \label{fig:GeoNeutrinoFlux}
\end{figure}

The parameters used in Eq.~\ref{equ:SurvivalProbability2} are listed in Ref.~\cite{PDG_2024}.
Since the neutrino-electron elastic scattering channel (Eq.~\ref{equ:ES}) is open to all flavors, the oscillation of electron (anti)neutrinos into $\nu_\mu$ and $\nu_\tau$ must be taken into account.
For the appearance probability of $\mu$ and $\tau$ neutrinos, it is
\begin{equation}
    \begin{aligned}
        P_{e\mu,e\tau}(E, L) = 1- P_{ee}(E, L).
    \end{aligned}
    \label{equ:SurvivalProbability3}
\end{equation}
Fig.~\ref{fig:GeoNeutrinoFlux} is the predicted electron (anti)neutrino spectrum at Jinping.
The results for other flavors are not shown.

\subsection{Elastic scattering event spectrum}
\begin{figure}
    \centering
    \includegraphics[width=\columnwidth]{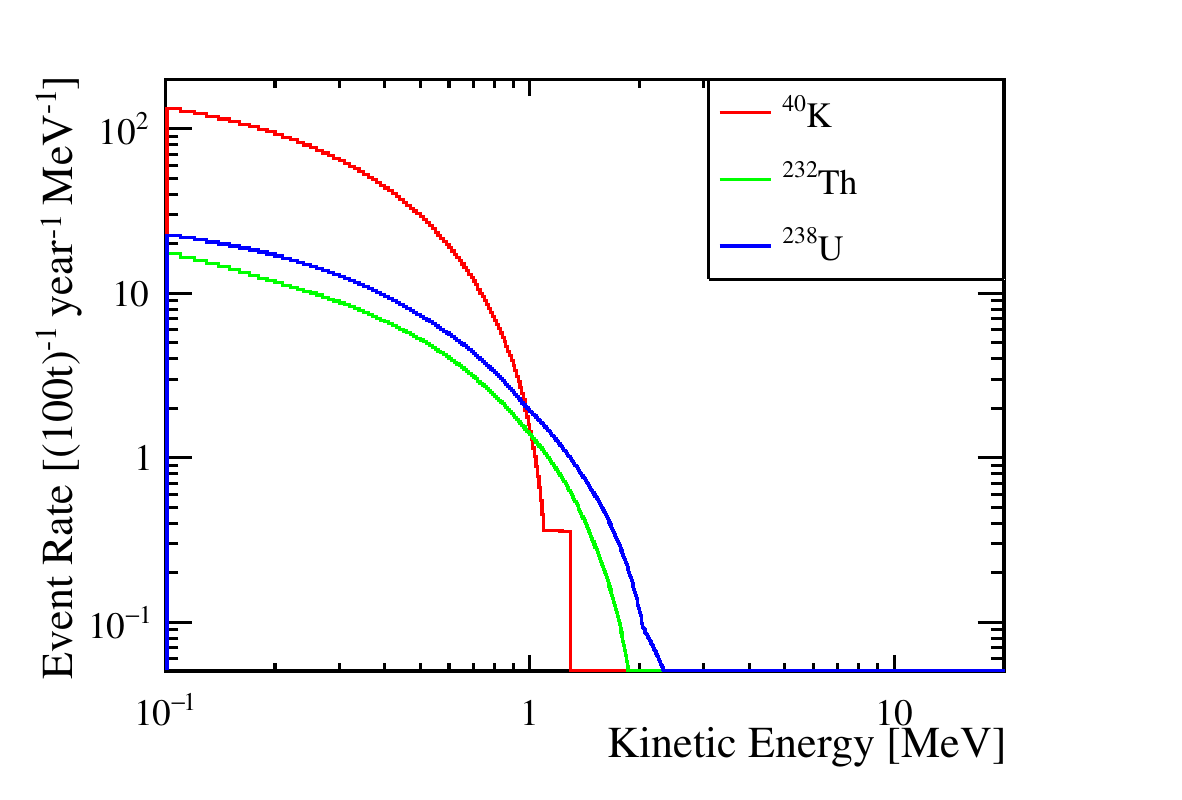}
    \caption{Total kinetic energy spectra of recoil electrons from U, Th, and K geoneutrinos at Jinping. The plot includes contributions from all flavors after neutrino oscillations.}
    \label{fig:GeoSignal1D}
\end{figure}

The differential cross-section ${\rm d} \sigma(E_\nu, T_e) / {\rm d} T_e$ for the neutrino-electron elastic scattering is from Ref.~\cite{CrossSection}, and the scattering event rate is obtained by
\begin{equation}
    \frac{{\rm d} n(T_e)}{{\rm d}T_e} = N_e \sum_k \int \frac{{\rm d} \sigma_k(E_\nu, T_e)}{{\rm d} T_e} \phi_k(E_\nu) {\rm d} E_\nu,
\end{equation}
where $T_e$ is the recoil electron kinetic energy, $N_e$ is the number of electrons of the detector target, and $k$ runs over all six neutrino types: $\nu_e$, $\nu_\mu$, $\nu_\tau$, $\bar{\nu}_e$, $\bar{\nu}_\mu$, and $\bar{\nu}_\tau$. The flux $\phi_k(E_\nu)$ for each flavor is obtained from the electron (anti)neutrino flux predictions combined with the flavor oscillation probabilities given in Eqs.~\ref{equ:SurvivalProbability2}--\ref{equ:SurvivalProbability3}.
The total kinetic energy spectra of all the recoil electrons from U, Th, and K are shown in Fig.~\ref{fig:GeoSignal1D}.

Additionally, the scattering angle, which is the deviation from the recoil electron's direction to the incident neutrino's direction, is a function of the neutrino energy $E_\nu$ and the recoil electron kinetic energy $T_e$:
\begin{equation}
    \label{eq:theta}
    \theta = \arccos \left( \frac{1 + m_e / E_\nu}{\sqrt{1 + 2m_e / T_e}} \right),
\end{equation}
which is used in the following directionality analysis. 

\subsection{Solar neutrino background}
\begin{figure}
    \centering
    \includegraphics[width=\columnwidth]{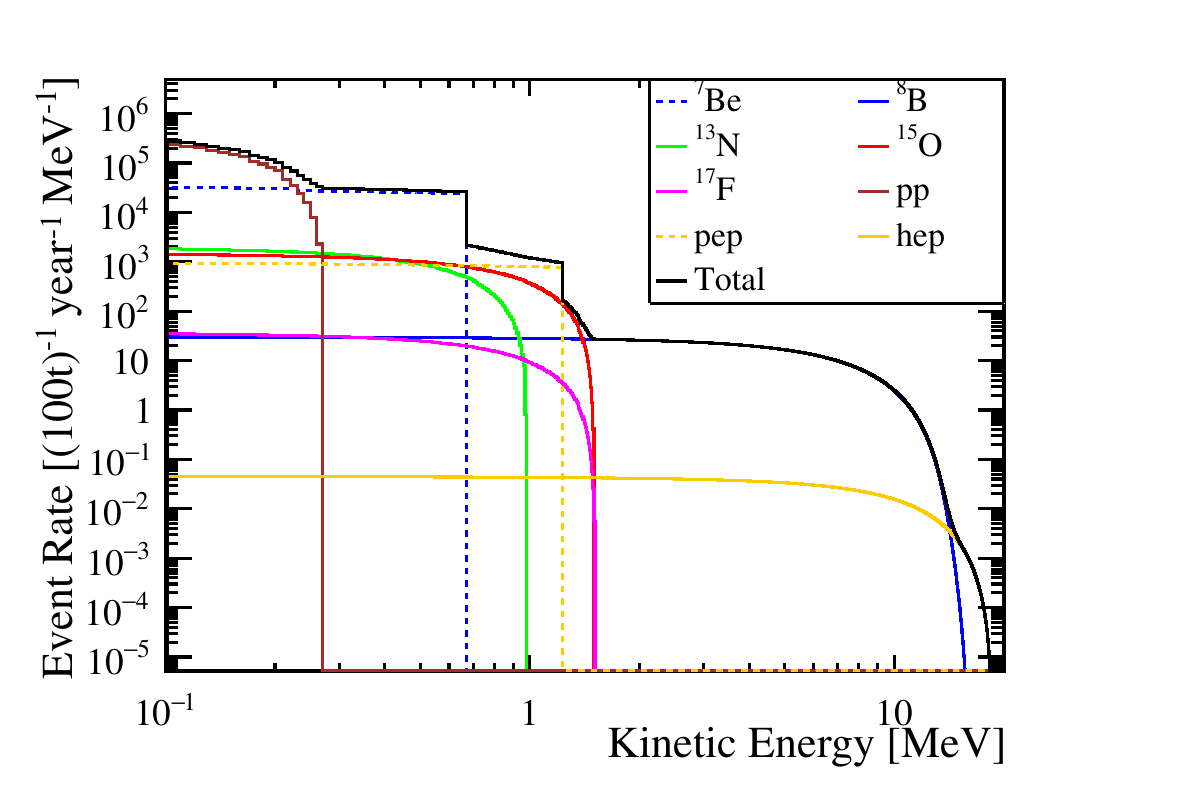}
    \caption{
    Total kinetic energy spectra of the recoil electrons from solar neutrinos at Jinping. The plot includes contributions from all flavors after neutrino oscillations.}
    \label{fig:SolarSignal1D_truth}
\end{figure}

The intrinsic background of the geoneutrino signal in neutrino-electron elastic scattering is from the solar neutrino. The solar neutrino flux and recoil electron spectrum are computed, as in the previous work~\cite{Wang_2020}, and the total recoil electron kinetic energy spectrum at Jinping is shown in Fig.~\ref{fig:SolarSignal1D_truth}.

\begin{figure}
    \centering
    \includegraphics[width=\columnwidth]{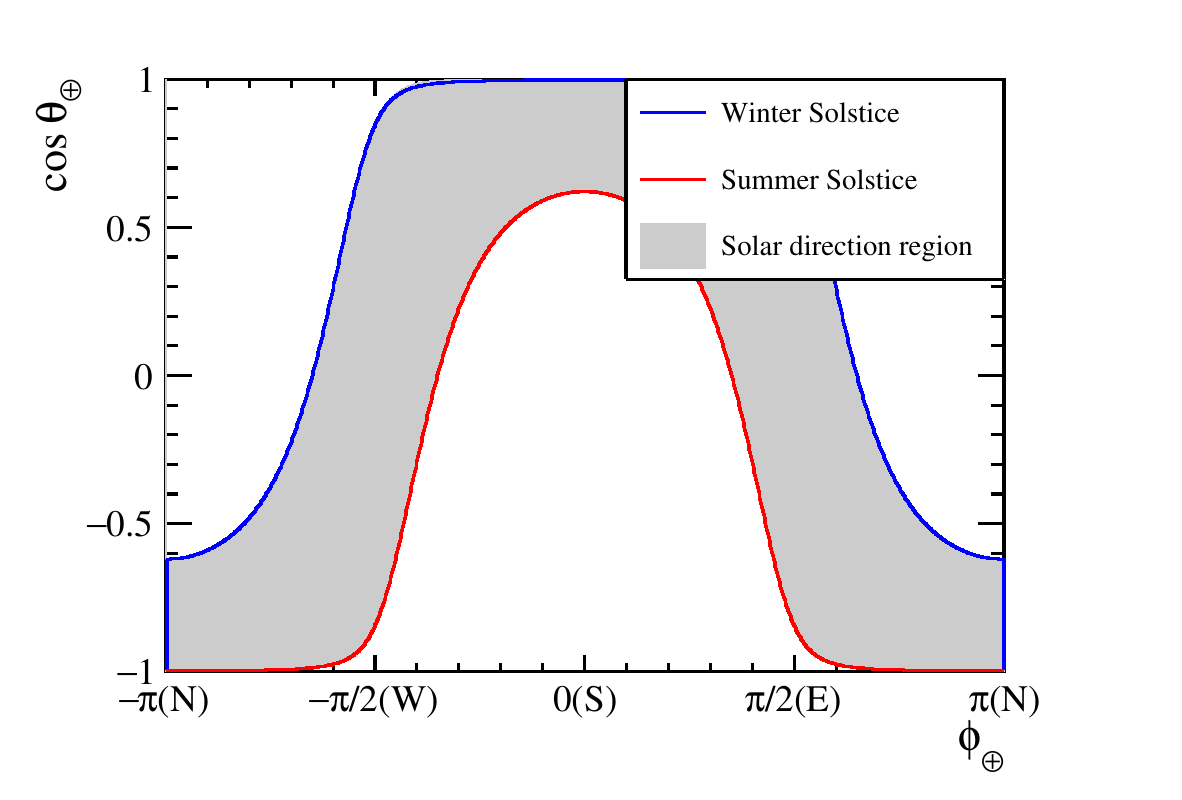}
    \caption{The evolution distribution of the terrestrial coordinate $\cos\theta_\oplus$ and  $\phi_{\oplus}$ of the solar neutrinos.
    The blue solid line represents the trajectory at the winter solstice, while the red solid line shows that at the summer solstice. The gray area represents the band for other times.}
    \label{fig:SolarDirection}
\end{figure}

The solar neutrino direction is the key to suppressing this background.
The Sun's direction varies periodically in the terrestrial coordinate system, exhibiting a minor daily cycle and a major annual cycle. 
As shown in Fig.~\ref{fig:SolarDirection},
the Sun's direction is plotted in the terrestrial coordinate system, where its daily trajectory forms a continuous curve spanning the azimuthal range from $\phi_\oplus = -\pi$ to $\phi_\oplus = \pi$.
The curves corresponding to the winter and summer solstices define the boundaries.
Solar neutrinos in this directional band can be parallel to the geoneutrinos, resulting in a substantial solar neutrino background.

\subsection{Internal radioactive background}
Radioactive isotopes contaminated in the liquid scintillator have to be taken into account. 
A Geant4-based simulation~\cite{Geant41,Geant42} is developed to simulate the full decay chains of $\rm ^{238}U$ and $\rm ^{232}Th$.
Both $\rm ^{238}U$ and $\rm ^{232}Th$ undergo a series of $\alpha$ and $\beta$ decays, ultimately reaching stable lead isotopes. 
The $\alpha$, $\beta$, and $\gamma$ particles emitted in these decays produce scintillation light and mimic the neutrino-electron elastic scattering signal. 

\begin{figure}
    \centering
    \begin{subfigure}{\columnwidth}
        \centering
        \includegraphics[width=\columnwidth]{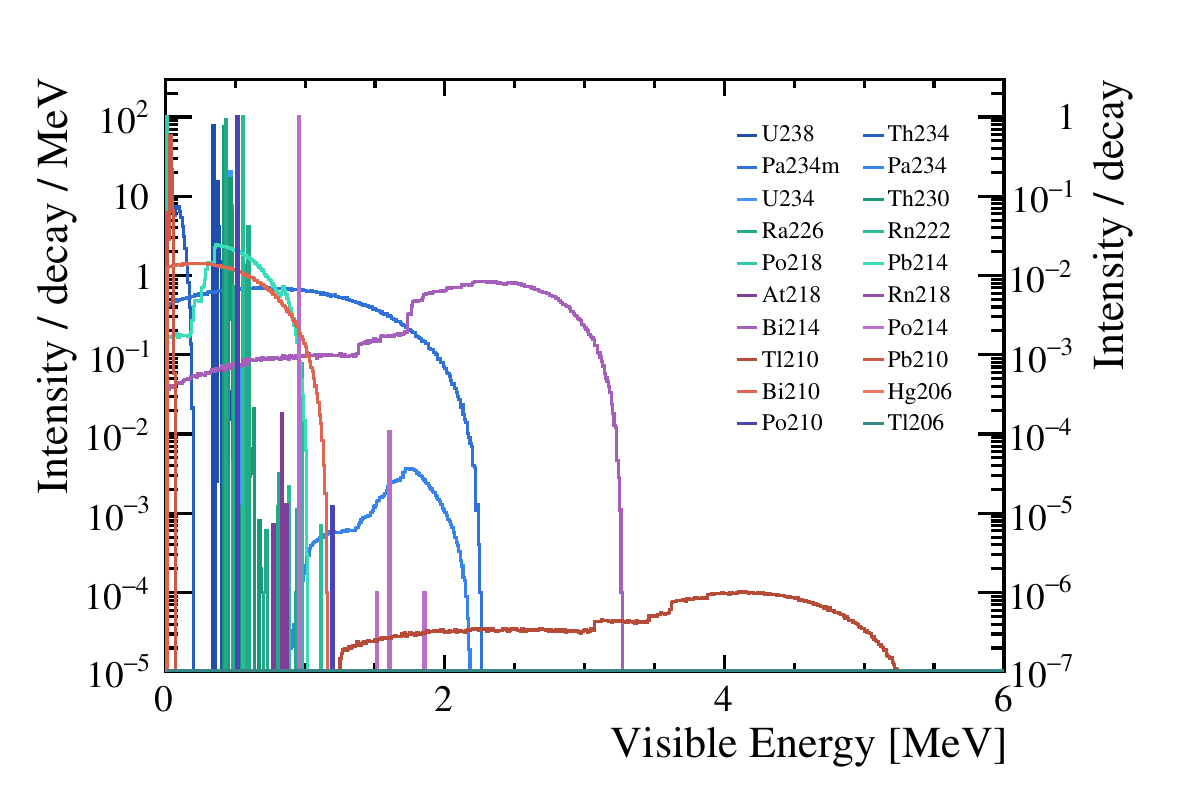}
        \caption{}
        \label{fig:RadioactiveSpectrum_U}
    \end{subfigure}
    \begin{subfigure}{\columnwidth}
        \centering
        \includegraphics[width=\columnwidth]{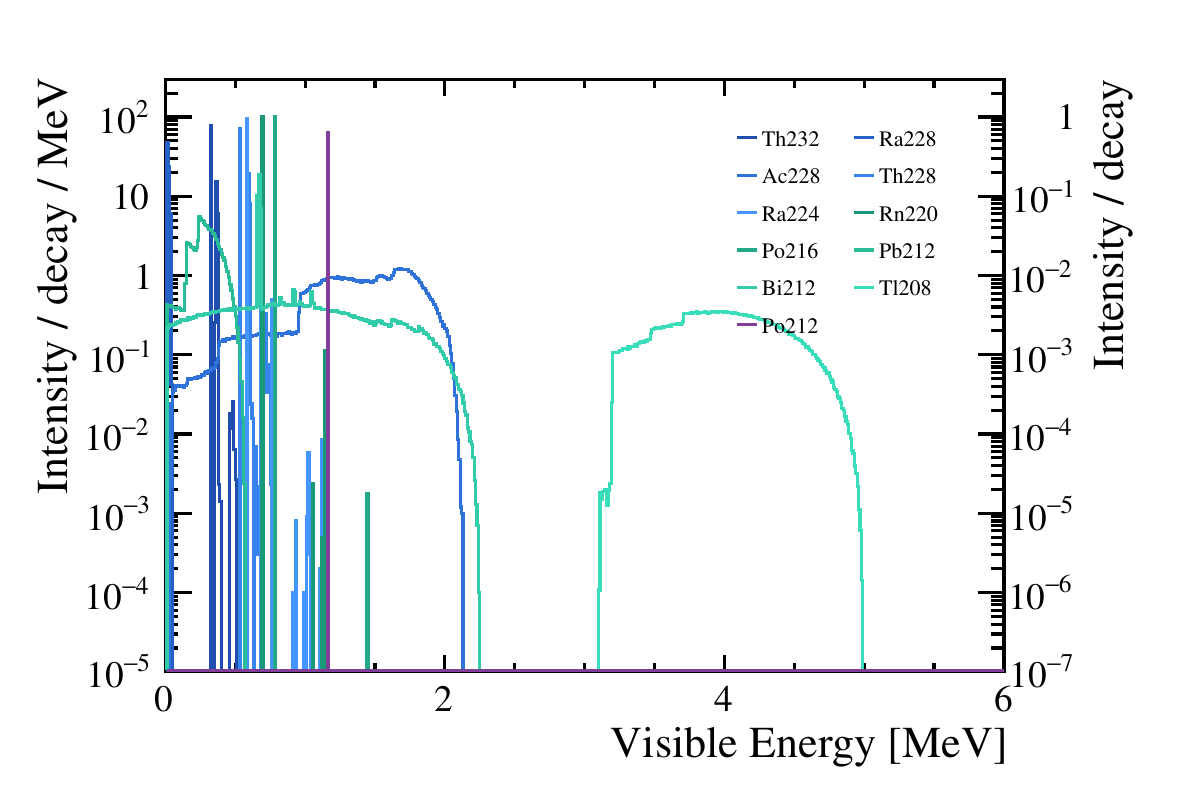}
        \caption{}
        \label{fig:RadioactiveSpectrum_Th}
    \end{subfigure}
    \caption{The visible energy spectra of the full $\rm ^{238}U$ decay chain and the full $\rm ^{232}Th$ decay chain, obtained from the Geant4 simulation. Quenching of $\alpha$ is taken into account.}
    \label{fig:RadioactiveSpectrum}
\end{figure}

Visible energy, $E_{\rm vis}$, is defined as the kinetic energy of $\beta$ and $\gamma$.
The $\alpha$ particles suffer from ionization quenching in liquid scintillators, which substantially reduces their light yield relative to electrons of the same energy. The relation between the visible energy $E_{\rm vis}$ and the true $\alpha$ particle energy $E_\alpha$ (in MeV) is given by the formula proposed by Ref.~\cite{von_2016}.
\begin{equation}
    E_{\rm vis}~{\rm [MeV]} =
    \begin{cases}
        0.031\,E_\alpha^{1.689}, & E_\alpha < 6.76~{\rm MeV}, \\[4pt]
        -0.505 + 0.190\,E_\alpha, & E_\alpha > 6.76~{\rm MeV}.
    \end{cases}
    \label{eq:Quenching}
\end{equation}
Owing to the short lifetimes of some metastable states,
successive decays within a few ns are grouped together. 
The visible energy spectra of the full $\rm ^{238}U$ decay chain and the full $\rm ^{232}Th$ decay chain, obtained from the Geant4 simulation and quenching simulation, are shown in Fig.~\ref{fig:RadioactiveSpectrum}. According to the Borexino experiment~\cite{Borexino_2020}, the contamination levels of $\rm ^{238}U$ and $\rm ^{232}Th$ are set to $9.4 \times 10^{-20}$ g/g and $5.7 \times 10^{-19}$ g/g, respectively.

\section{Detector response simulation}
\label{sec:Simulation}

\subsection{Detector setup and simulation}
\label{sec:simulation}
We adopted the detector scheme used in Ref.~\cite{Wang_2020}, a popular arrangement with liquid scintillators, as in the KamLAND, Borexino, and SNO+ experiments. The center of the detector is the main liquid scintillator, contained in a transparent sphere (i.e.~an acrylic shell). Surrounding the sphere are the photomultiplier tubes (PMTs), which collect the optical photons emitted during the Cherenkov and scintillation processes of charged particles.
In the Geant4 simulation~\cite{Geant41, Geant42}, the electromagnetic and optical processes are turned on, incident electrons with kinetic energy from 0 to 3~MeV are produced in the liquid scintillator sphere, and the hit time of each photoelectron is recorded for subsequent reconstruction.

\subsection{Cherenkov liquid scintillator}
\begin{figure}
    \centering
    \includegraphics[width=\columnwidth]{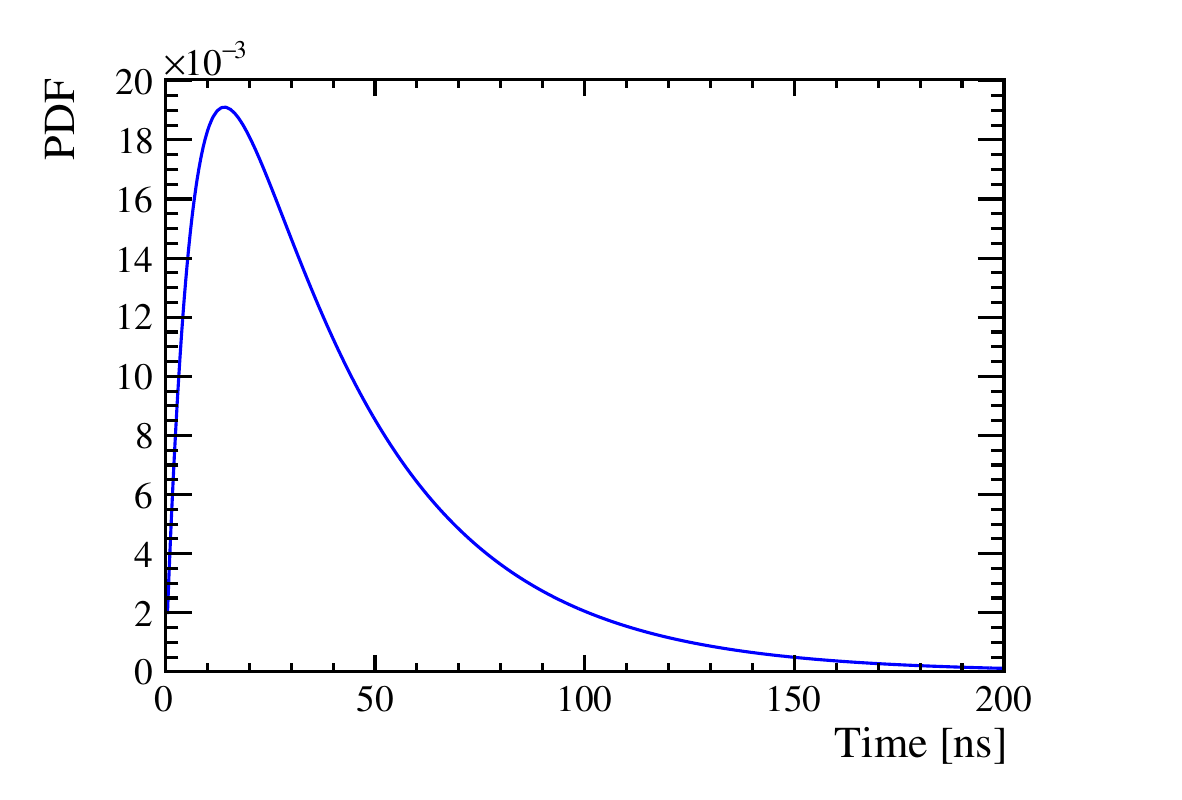}
    \caption{Scintillation light emission curve of the Cherenkov liquid scintillator.}
    \label{fig:LightCurve}
\end{figure}

There are several Cherenkov liquid scintillator candidates, which all have a long scintillation emission time and a reasonable light yield~\cite{Li_2016, Guo_2017, Steven_2020, Steiger}. 
In this research, we chose the one with a rise time of 7~ns. The light curve of the liquid scintillator is shown in Fig.~\ref{fig:LightCurve}.

The recoil electron scattering hinders the critical technical performance of directionality measurement in the liquid scintillator. 
The choice of a particular Cherenkov liquid scintillator is not a fundamental factor, because they all have similar density and element composition~\cite{Li_2016, Guo_2017, Steven_2020, Steiger}, and the capability for direction reconstruction~\cite{Luo_2023} is stable in a certain range. 

\subsection{Detector response}

In this work, we use a PMT's photocathode coverage rate of 50\% and a quantum efficiency of about 30\% for visible photons (350 nm to 800 nm). So we can collect 15\% of the produced photons to reconstruct the energy and direction of the incident particle. We choose the photons produced in the first 2 ns, which are expected to be Cherenkov photons mostly, to reconstruct the direction of the primary electron, $\vec{r}_{\rm rec}$, as shown in Eq.~\ref{equ:DirectionReconstruction} 
\begin{equation}
    \vec{r}_{\rm rec} = \frac{1}{N_p} \sum_{i = 1}^{N_p} \vec{r}_i,
    \label{equ:DirectionReconstruction}
\end{equation}
in which the $\vec{r}_i$ stands for the direction of the $i$-th photon, and $N_p$ represents the total number of selected photons.

\begin{figure}
    \centering
    \includegraphics[width=\columnwidth]{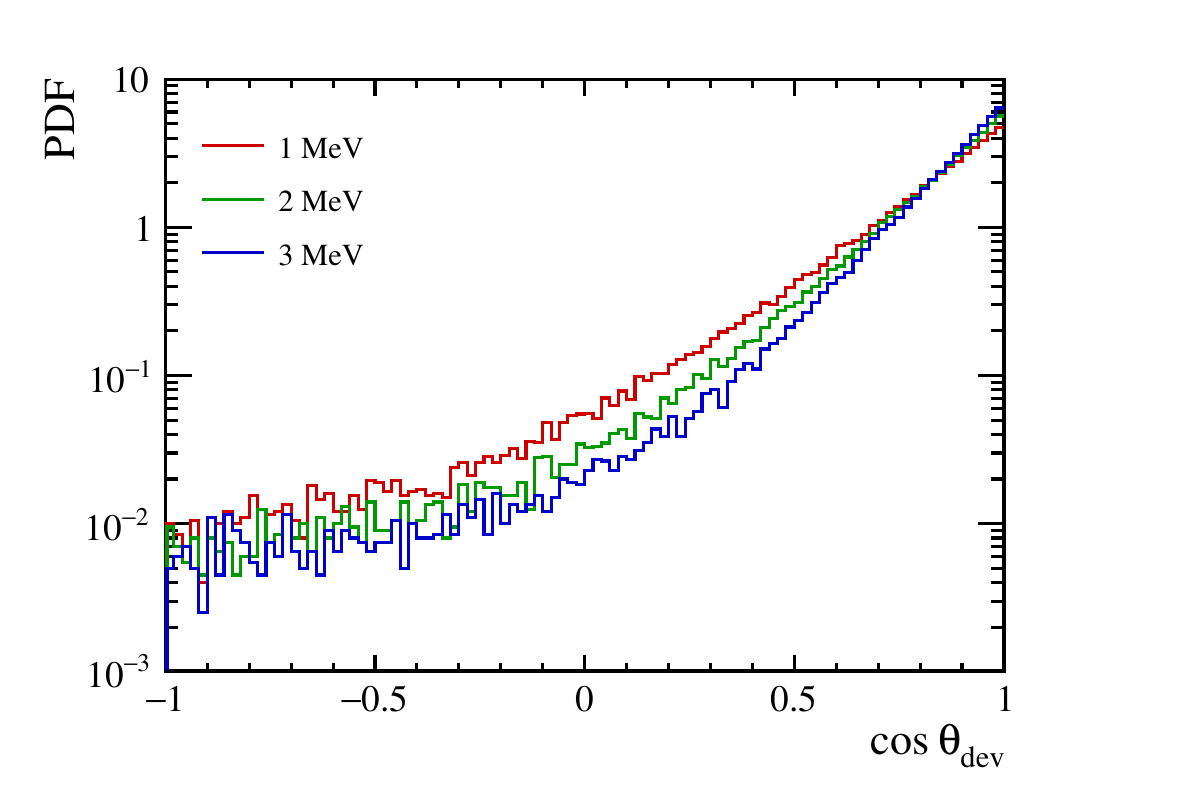}
    \caption{Distributions of the cosine of the angle between the reconstructed direction and the true recoil electron direction for 1, 2, and 3~MeV electrons.}
    \label{fig:AngularResponse}
\end{figure}

\begin{figure}
    \centering
    \includegraphics[width=\columnwidth]{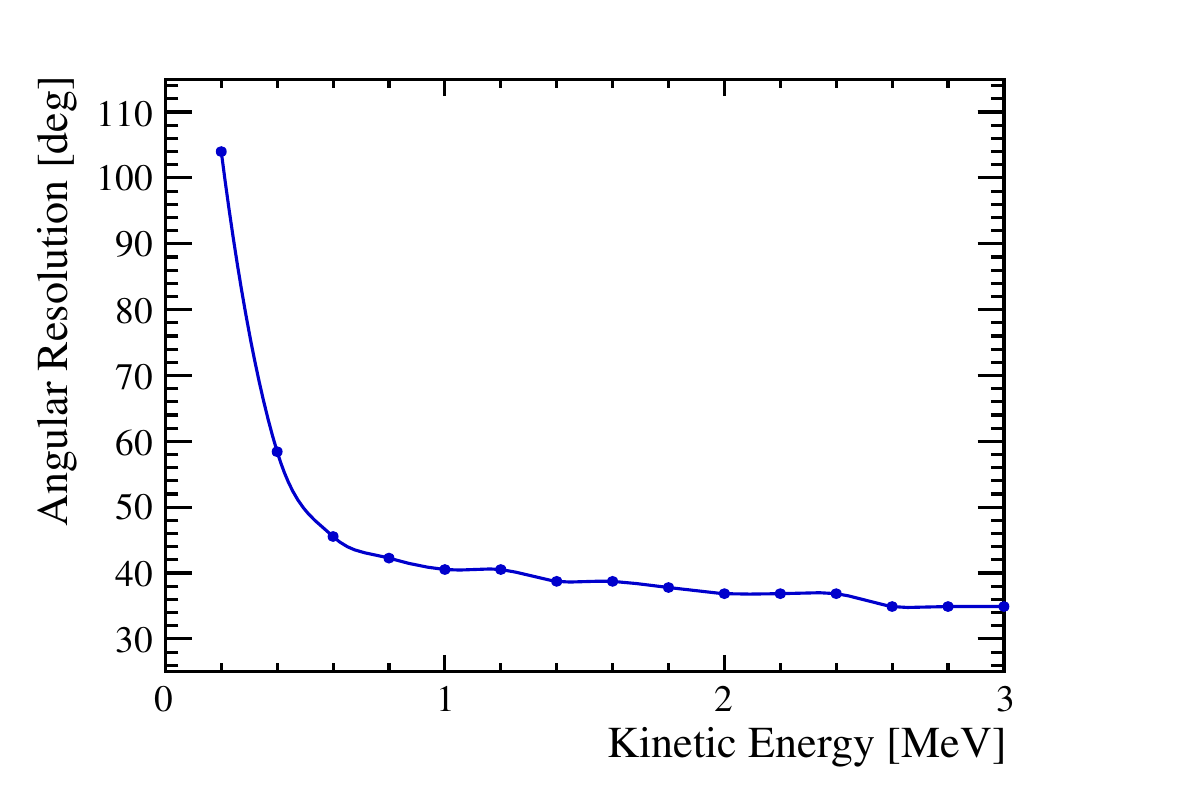}
    \caption{Angular resolution as a function of incident electron kinetic energy. The angular reconstruction performance deteriorates as the electron kinetic energy decreases.}
    \label{fig:AngularResolution}
\end{figure}

The cosine of the angle between the reconstructed direction and the true recoil electron direction is measured. The distributions for 1, 2, and 3 MeV are plotted in Fig.~\ref{fig:AngularResponse}.
The $1\sigma$ angular resolution (defined as the range enclosing 68\% of events) 
is plotted as a function of the kinetic energy of recoil electrons in Fig.~\ref{fig:AngularResolution}.
 
In the previous study~\cite{Wang_2020}, the number of protons of carbon was set to 12 in the Geant4 simulation, resulting in a much worse electron direction resolution than in reality. The $1\sigma$ angular resolution is 40 degrees, which was 55 degrees for a 1 MeV electron.

The particle identification capability of Cherenkov liquid scintillators has been preliminarily studied in previous work~\cite{Luo_2023}. In the present study, we adopt a relatively optimistic estimation: we assume that all $\alpha$-type radioactive background events can be rejected with negligible loss of the signal efficiency.

\section{Selection criteria}
\label{sec:Criteria}

In this section, the selection criteria for separating the geoneutrino signal from the backgrounds are established. First, three energy regions of interest ($\rm RoI_K$, $\rm RoI_{U/Th}$, and $\rm RoI_{Geo}$) are defined based on the signal-to-background ratio to target geoneutrinos from different elements. Second, a solar angle cut is applied and optimized to suppress the solar neutrino background while preserving the geoneutrino signal.

\subsection{Geo-solar ratio and energy region selection}

\begin{figure}
    \centering
    \includegraphics[width=\columnwidth]{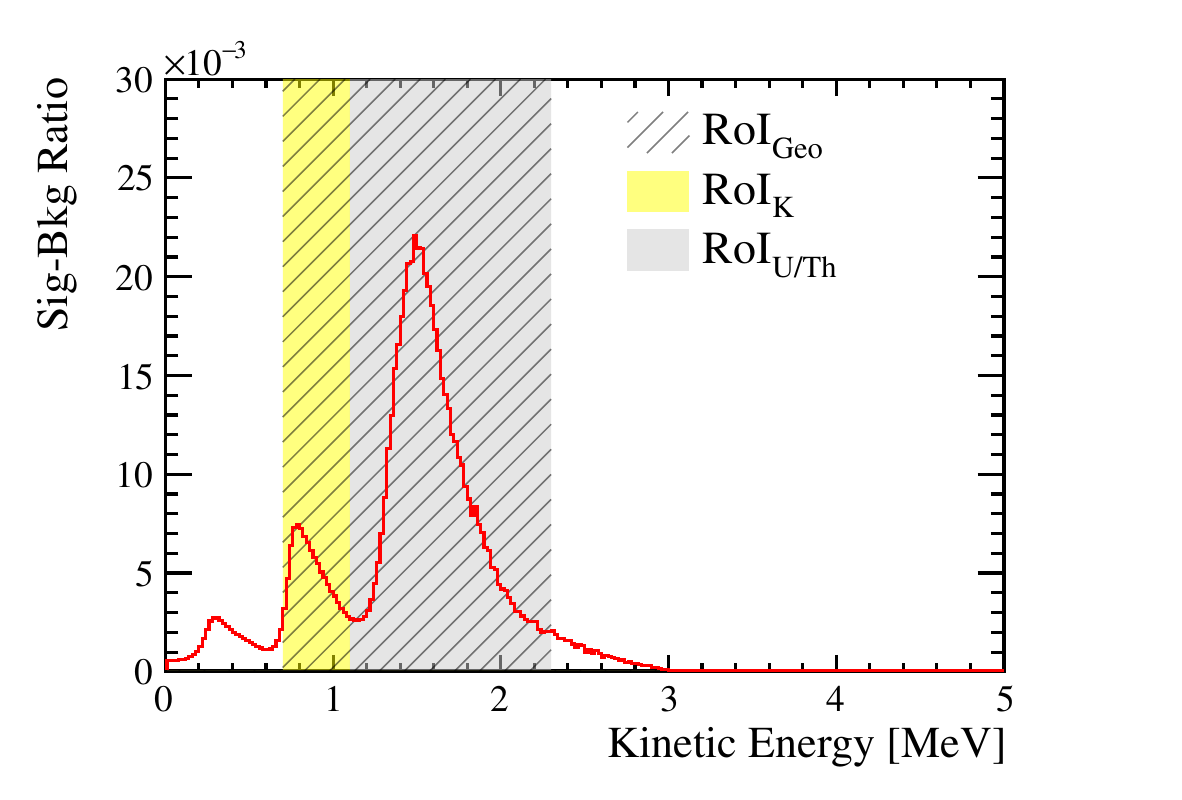}
    \caption{The ratio of the geoneutrino event rate to that of solar neutrino and radioactive background versus the recoil electron kinetic energy.}
    \label{fig:SignalRatio}
\end{figure}

Three energy regions are selected for the research of geoneutrinos from different elements based on the signal-to-background ratio, as shown in Fig.~\ref{fig:SignalRatio}. The first region of interest, $\rm RoI_K$, covers an energy range of 0.7~MeV to 1.1~MeV, in which the geoneutrinos from $\rm ^{40}K$ are dominant. Similarly, $\rm RoI_{U/Th}$ covers an energy region of 1.1~MeV to 2.3~MeV, and is dominated by $\rm ^{238}U$ and $\rm ^{232}Th$ geoneutrinos. $\rm RoI_{Geo}$ covers an energy region of 0.7~MeV to 2.3~MeV, which contains almost all the geoneutrino signals. In $\rm RoI_K$ and $\rm RoI_{U/Th}$, the dominant background components are $\rm ^7Be$, pep, and $\rm ^{15}O$ neutrinos, while the main backgrounds are pep, $\rm ^{15}O$, and $\rm ^8B$ neutrinos in $\rm RoI_{U/Th}$.

\subsection{Neutrino signal direction and solar angle cut}

\begin{figure}
    \centering
    \includegraphics[width=\columnwidth]{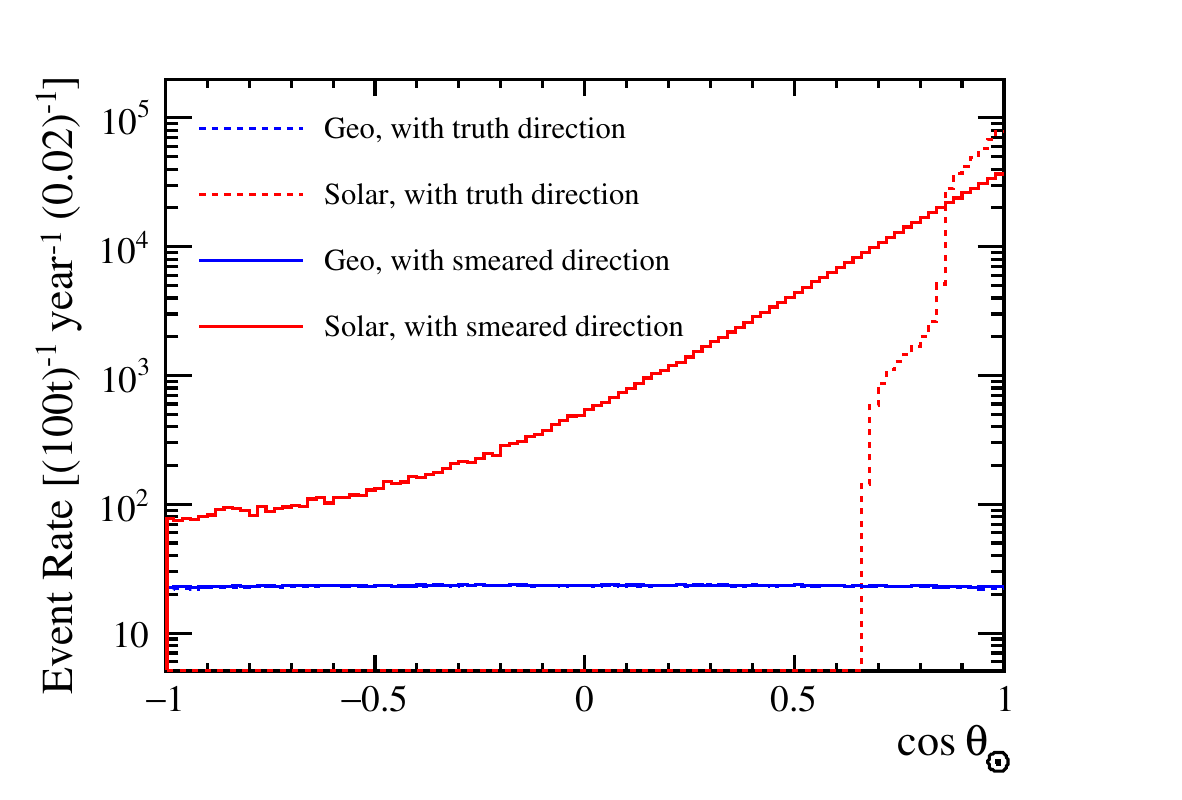}
    \caption{The solar angle distribution for the geoneutrino signals and solar-neutrino backgrounds. Blue dashed and solid lines represent the distribution of geoneutrino signals with the recoiling electron's kinetic energy in $\rm RoI_{Geo}$, with true direction and smeared direction, respectively. Similarly, red lines represent the same distributions for solar-neutrino signals. The situation is similar for $\rm RoI_{K}$ and $\rm RoI_{U/Th}$.}
    \label{fig:SolarAngle}
\end{figure}

\begin{figure*}
    \centering
    \begin{subfigure}{\columnwidth}
        \includegraphics[width=\columnwidth]{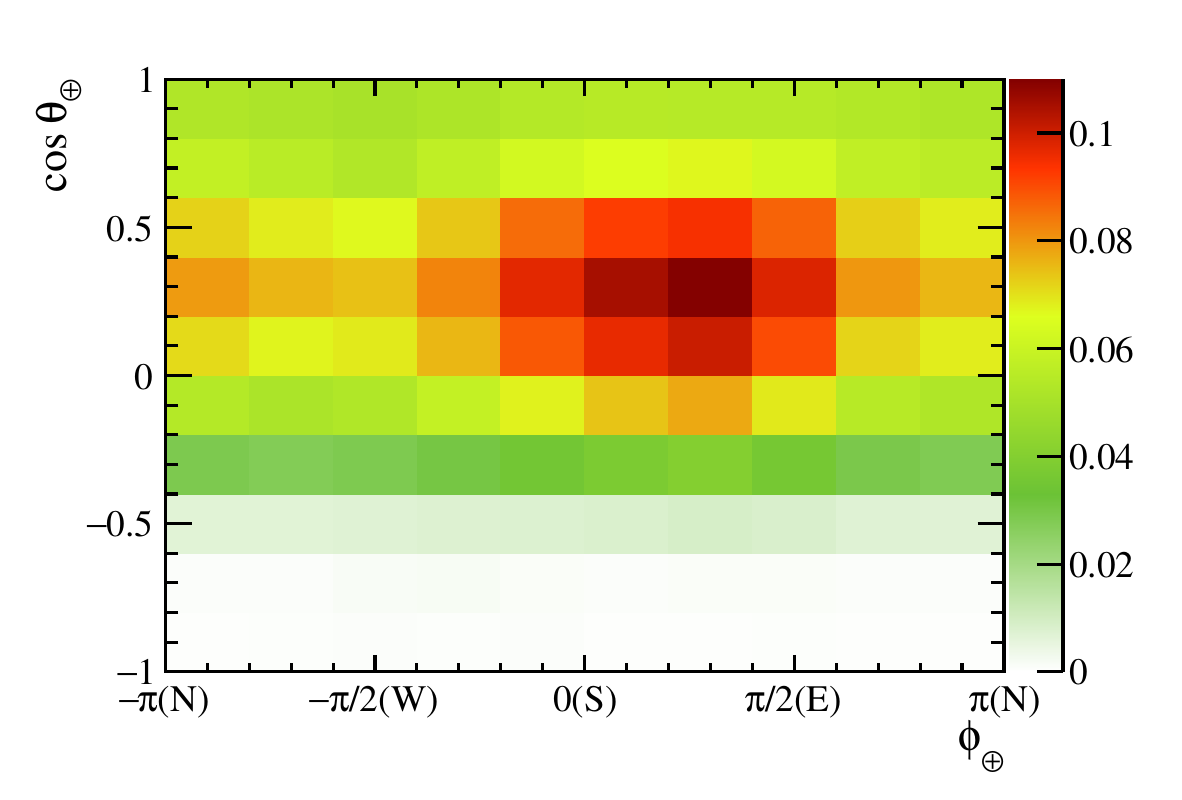}
        \caption{}
        \label{fig:GeoDirection_truth}
    \end{subfigure}
    \begin{subfigure}{\columnwidth}
        \includegraphics[width=\columnwidth]{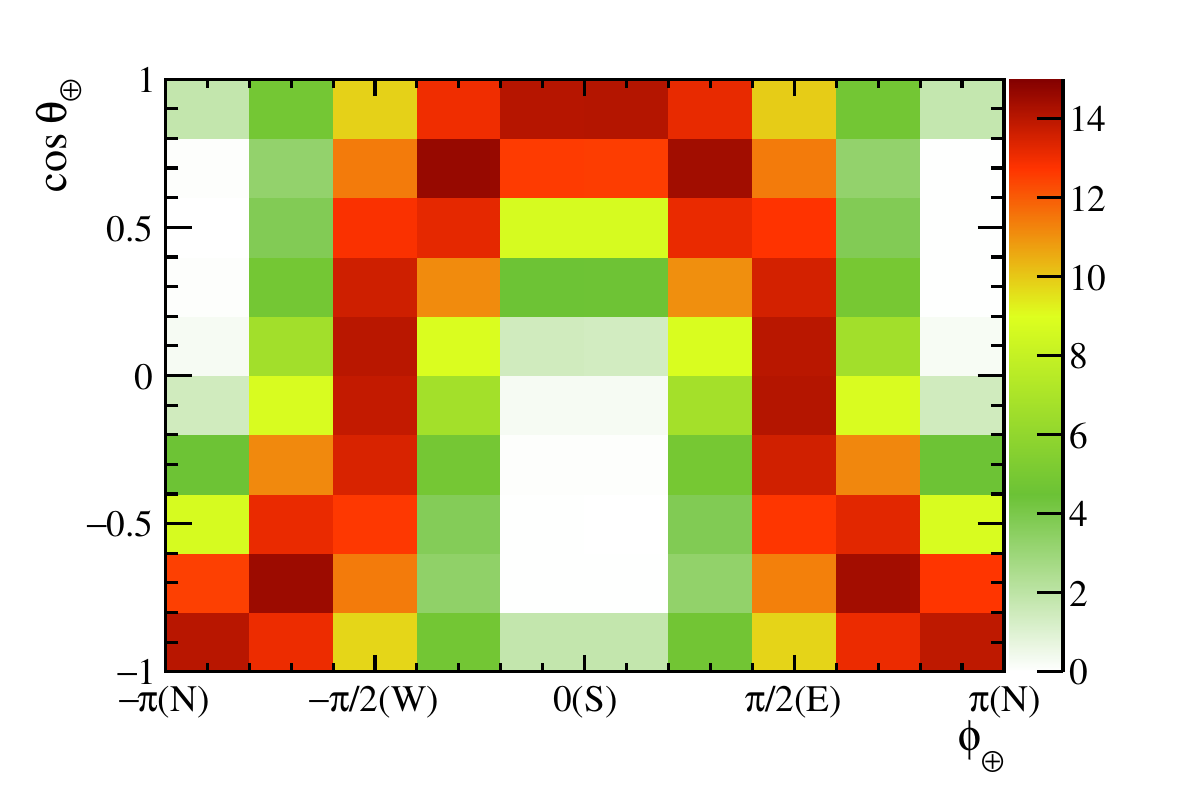}
        \caption{}
        \label{fig:SolarDirection_truth}
    \end{subfigure}
    \begin{subfigure}{\columnwidth}
        \includegraphics[width=\columnwidth]{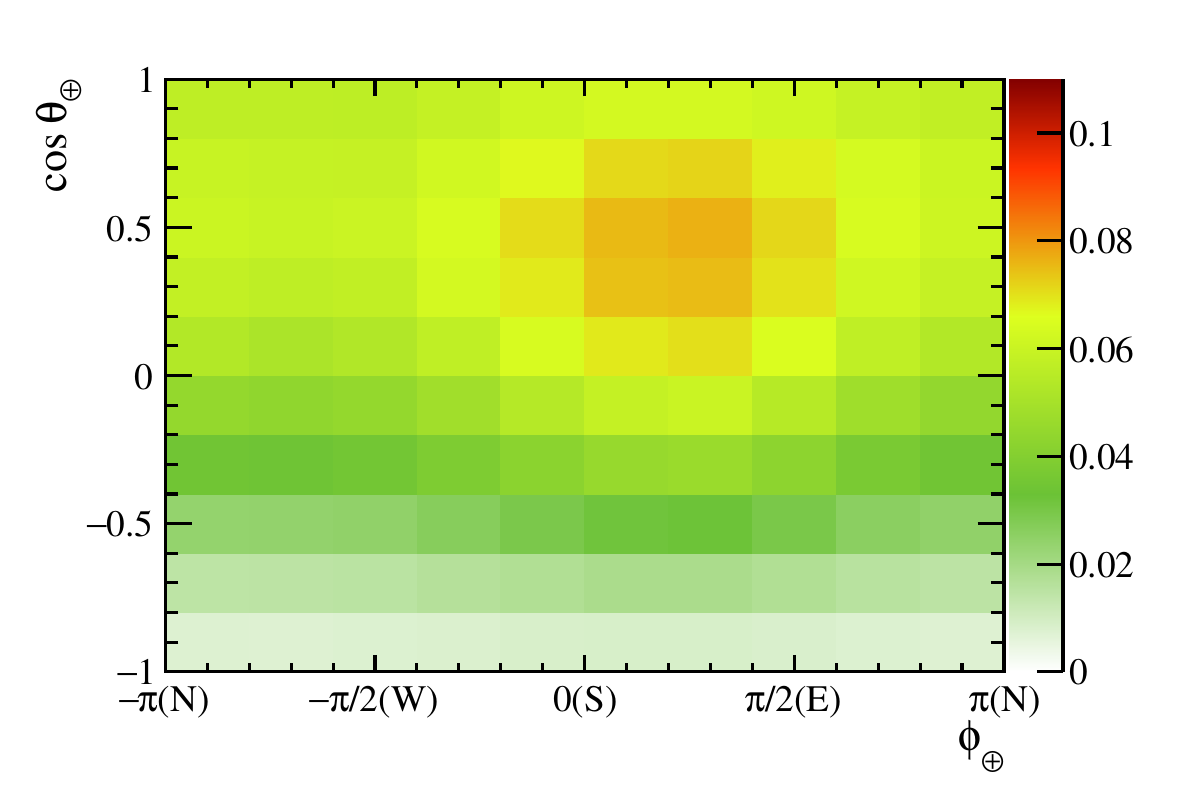}
        \caption{}
        \label{fig:GeoDirection_recon}
    \end{subfigure}
    \begin{subfigure}{\columnwidth}
        \includegraphics[width=\columnwidth]{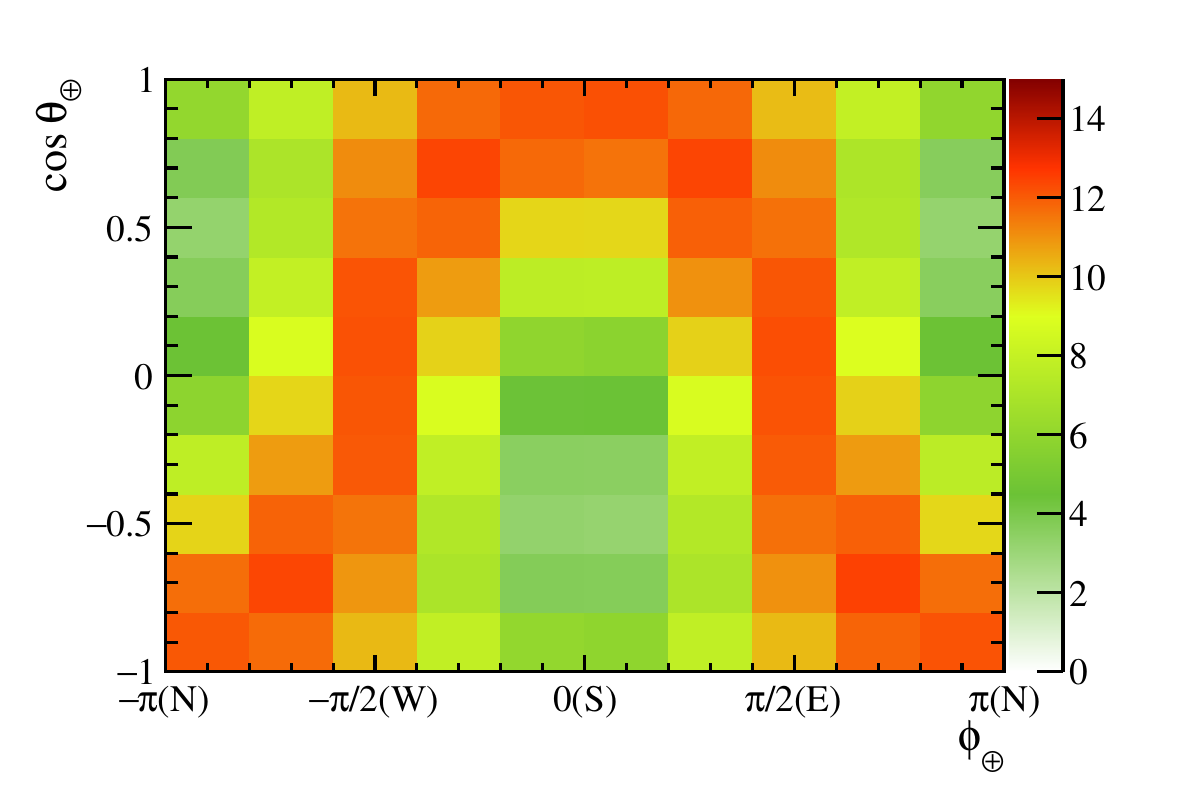}
        \caption{}
        \label{fig:SolarDirection_recon}
    \end{subfigure}
    \caption{Angular distributions of the recoil electrons produced by the geoneutrinos and solar neutrinos in $\rm RoI_{Geo}$. Panels (a) and (b) show the true angular distributions for geoneutrinos and solar neutrinos, respectively, while panels (c) and (d) show the corresponding reconstructed angular distributions for electrons. The $z$-axis of each panel shows the expected statistics of 100~ton-years.}
    \label{fig:NeutrinoDirection}
\end{figure*}

Most of the detected geoneutrinos are emitted from the near-field sediments and upper crust, and some are from the mantle.
Due to the rotation and revolution of the Earth, the direction of geoneutrinos has almost no correlation with the Sun's direction.
The angular distributions in solar angle of the geoneutrinos and the reconstructed electrons are both uniform, as presented in 
Fig.~\ref{fig:SolarAngle} (dashed and solid blue lines).
On the other hand, in the terrestrial coordinate system, geoneutrinos gather at $\cos \theta_\oplus>0$.
They also show a cluster structure at $0 < \phi_\oplus < \pi / 2$, manifesting the relatively thicker crust in the Qinghai-Tibet Plateau, which is located northwest of the detector. 
Figure~\ref{fig:GeoDirection_truth} and Fig.~\ref{fig:GeoDirection_recon} show the $\cos\theta_\oplus$ and  $\phi_{\oplus}$
distributions for the geoneutrinos and reconstructed electrons, respectively.

Solar neutrinos arrive at the Earth in parallel, and their direction depends on the Sun's position at each moment.
After the neutrino-electron elastic scattering, 
the recoil electron will deviate from the true neutrino direction, but after the low-energy cut, i.e., the 0.7 MeV cut, the consistency between the
two directions is recovered (see Eq.~\ref{eq:theta} and the red dashed line of Fig.~\ref{fig:SolarAngle} for $\rm RoI_{Geo}$).
These electrons are scattered in the Cherenkov liquid scintillator, detected with PMTs, and finally reconstructed. 
The scattering, detection, and reconstruction processes further smear the electron directions and deviate them from the Sun's direction (see the red solid line in Fig.~\ref{fig:SolarAngle} for $\rm RoI_{Geo}$.
If observed in the terrestrial coordinate system, the $\cos\theta_\oplus$ and  $\phi_{\oplus}$ of these solar neutrinos have the distribution as in Fig.~\ref{fig:SolarDirection_truth}.
The distribution of $\cos\theta_\oplus$ and  $\phi_{\oplus}$ of reconstructed electrons is shown in Fig.~\ref{fig:SolarDirection_recon}.

\begin{figure*}
    \centering
    \begin{subfigure}{\columnwidth}
        \includegraphics[width=\columnwidth]{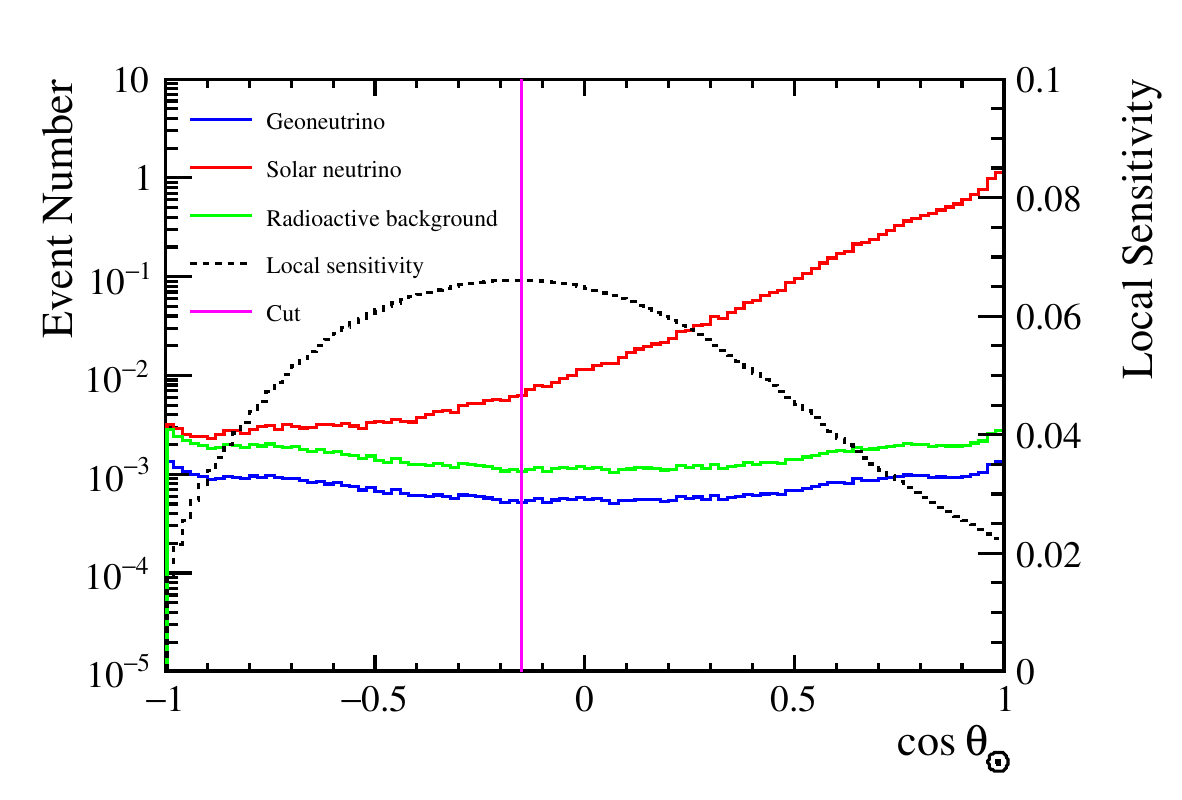}
    \end{subfigure}
    \begin{subfigure}{\columnwidth}
        \includegraphics[width=\columnwidth]{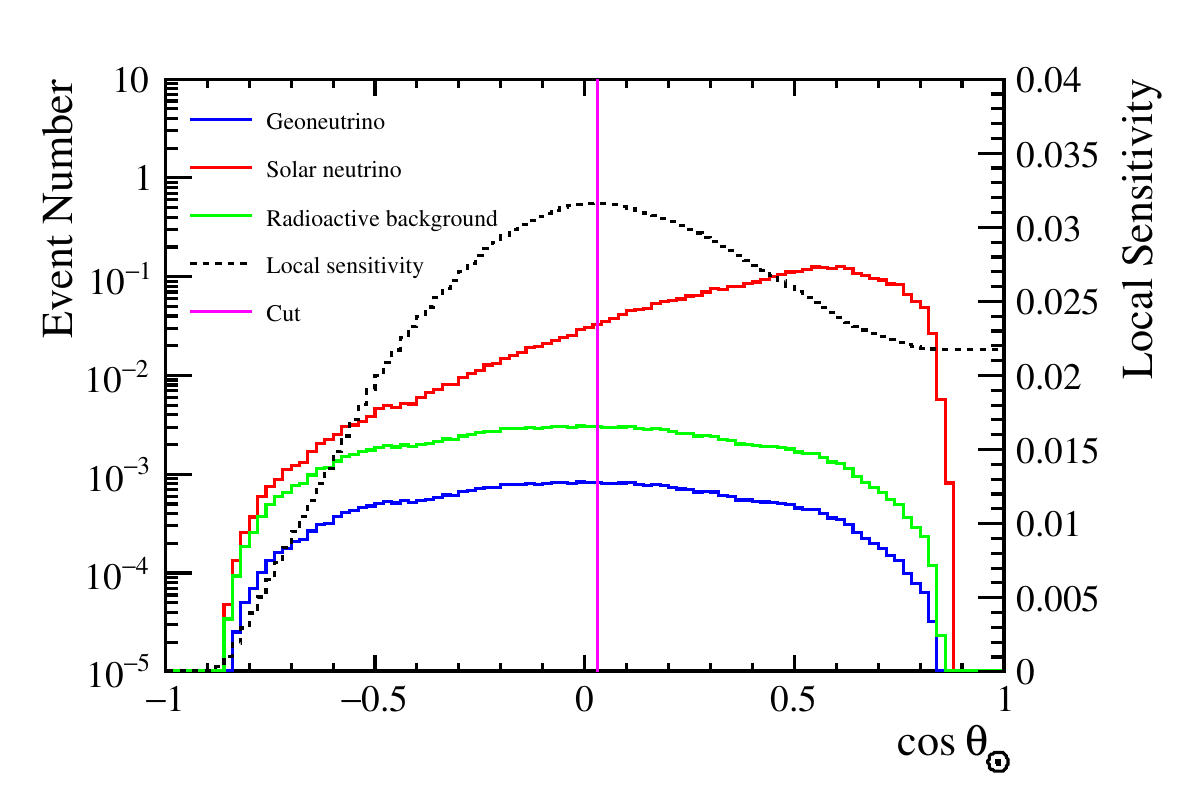}
    \end{subfigure}
    \caption{The $\cos \theta_\odot$ cut is optimized for each terrestrial solid angle cell. 
    The first figure represents the result for events with reconstructed directions of recoil electrons in the terrestrial solid angle cell of $\cos \theta_\oplus \in (0.4, 0.6)$ and $\phi_\oplus \in (0.4 \pi, 0.6 \pi)$, 
    while the second one is for the cell of $\cos \theta_\oplus \in (-0.6, -0.4)$ and $\phi_\oplus \in (-0.2 \pi, 0)$.
    The blue, red, and green solid lines show the $\cos \theta_\odot$ distribution for the reconstructed directions of the geoneutrino signals, solar neutrino backgrounds, and radioactive backgrounds under an exposure of 100~ton-years, respectively.
    The black dashed line shows the local sensitivity scan curve when changing the $\cos \theta_\odot$ cut as in Eq.~\ref{eq:localSens}, and the magenta line marks the maximum of the local sensitivity and the selected cut.
    In the second plot, the red, green, and blue curves cannot reach $\pm1$, because the direction of the Sun's direct rays cannot go beyond the tropics.
    }
    \label{fig:CutScan}
\end{figure*}

\begin{figure}
    \centering
    \includegraphics[width=\columnwidth]{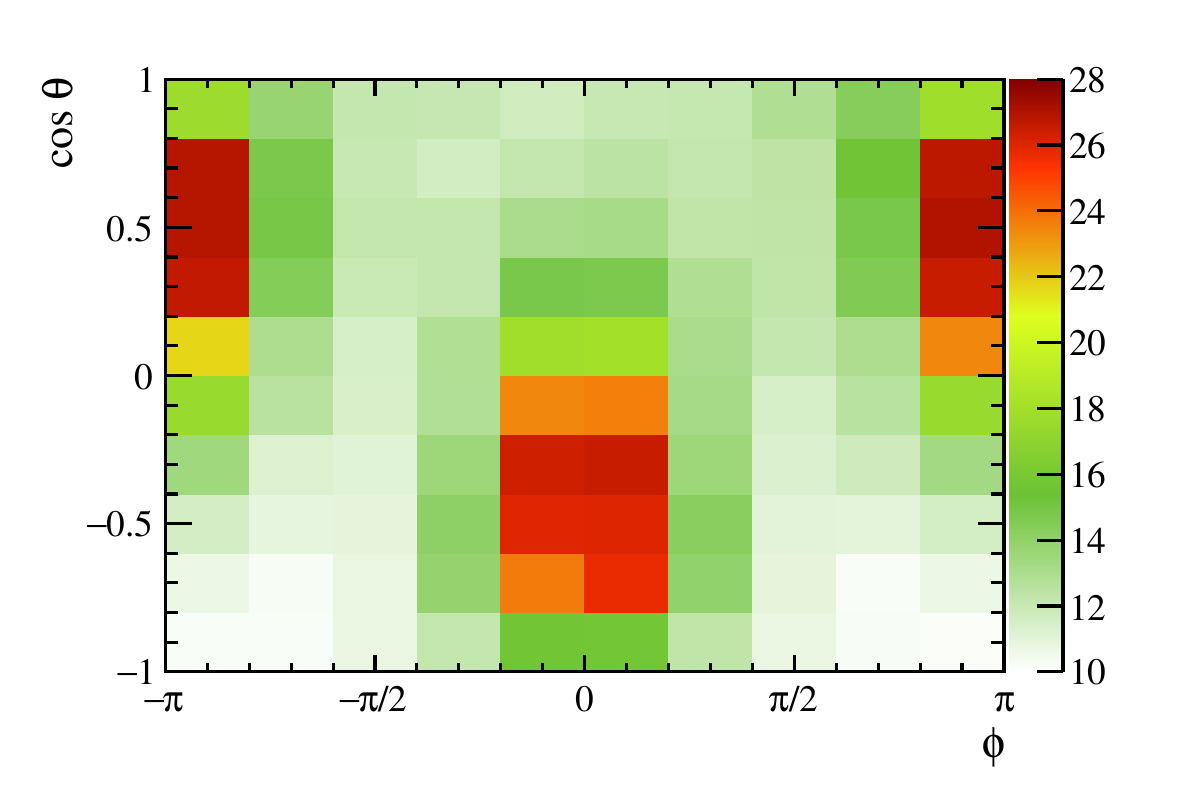}
    \caption{The terrestrial angle distribution of all expected candidates, including geoneutrinos, solar neutrinos, and radioactive backgrounds, after the $\cos \theta_\odot$ cut in $\rm RoI_{Geo}$ at an exposure of 5~kiloton-years. 
    The places where the Sun can't shine directly have the best signal-to-background ratio and the highest signal statistics.}
    \label{fig:SignalDirection2D}
\end{figure}

To fully utilize the directionality of the geoneutrino signal and to conduct subsequent analysis, the full $4 \pi$ space in the terrestrial coordinate system is divided into $10 \times 10$ solid-angle cells. The solar angle cut is optimized for each solid-angle cell. For a certain cell $i$, a critical cut boundary $\cos \theta_{\rm max}$ is chosen, and a solar angle cut $\cos \theta_\odot < \cos \theta_{\rm max}$ is applied to all candidates to maximize the local statistical sensitivity, $S_i^{\rm loc}$, 
\begin{equation}
    S_i^{\rm loc} = N_i^{\rm geo} / \sqrt{N_i^{\rm geo} + N_i^{\rm solar} + N_i^{\rm rad}},
    \label{eq:localSens}
\end{equation}
where $N_i^{\rm geo}$, $N_i^{\rm solar}$, and $N_i^{\rm rad}$ are the numbers of events of the geoneutrino signal, the solar neutrino background, and the radioactive background in the cell $i$, respectively.
Different terrestrial angle cells have different relative orientations to the Earth's center and the Sun, and the solar angle cuts differ accordingly. The situations for two example cells are shown in Fig.~\ref{fig:CutScan}. 
The first plot represents the optimization result for a typical solid angle cell covered by the Sun's trajectory, with $\cos \theta_\oplus \in (0.4, 0.6)$ and $\phi_\oplus \in (0.4 \pi, 0.6 \pi)$, which is located in the gray band in Fig.~\ref{fig:SolarDirection}. 
The second plot shows another typical solid angle cell outside the band with $\cos \theta_\oplus \in (-0.6, -0.4)$ and $\phi_\oplus \in (-0.2 \pi, 0)$.
The angular distribution of all candidates, including geoneutrinos and solar neutrinos, after the energy and $\cos \theta_\odot$ cuts are shown in Fig.~\ref{fig:SignalDirection2D}, in which the reconstructed energy is in $\rm RoI_{Geo}$, and the exposure is 5~kiloton-years. 
The places where the Sun can't shine directly have the best signal-to-background ratio and the highest signal statistics.
The event rates before and after the $\cos \theta_\odot$ cut in the three energy regions of interest are listed in Tab.~\ref{tab:EventRate}.

\begin{table*}
    \centering
    \caption{Number of events for each geoneutrino component at an exposure of 5 kiloton-years, with values shown for both before and after the $\cos \theta_\odot$ cut.}
    \begin{tabular}{ccc|cc|cc}
        \toprule
        \toprule
        &\multicolumn{2}{c|}{$\rm RoI_{Geo}$}    &\multicolumn{2}{c|}{$\rm RoI_{K}$}    &\multicolumn{2}{c}{$\rm RoI_{U/Th}$}    \\
        &Before $\cos \theta_\odot$ cut &After $\cos \theta_\odot$ cut    &Before $\cos \theta_\odot$ cut &After $\cos \theta_\odot$ cut    &Before $\cos \theta_\odot$ cut &After $\cos \theta_\odot$ cut    \\
        \midrule
        $\rm ^{40}K$ &95.5  &44.1  &91.9  &39.8  &3.6  &2.0  \\
        $\rm ^{232}Th$ &54.3  &25.1  &38.4  &16.6  &15.9  &8.7  \\
        $\rm ^{238}U$ &83.0  &38.3  &53.3  &23.1  &29.7  &16.2  \\
        Geoneutrino &232.8  &107.5  &183.6  &79.5  &49.2  &26.9  \\
        Solar neutrino &44206.8  &1024.2  &35476.2  &748.4  &8530.5  &235.2  \\
        Radioactive bkg. & 778.1 & 357.5 & 259.9 & 111.7 & 518.1 & 282.2 \\
        \bottomrule
        \bottomrule
    \end{tabular}
    \label{tab:EventRate}
\end{table*}

\section{Sensitivity calculation}
\label{sec:Sensitivity}

\subsection{Geoneutrino detection sensitivity}
\label{sec:GeoSens}

Two methods are applied to calculate the discovery sensitivity in three energy regions of interest ($\rm RoI_K$, $\rm RoI_{U/Th}$, and $\rm RoI_{Geo}$).

Firstly, a total-event-rate method is used to estimate the detection sensitivity. The total numbers of events of the geoneutrino signal and solar neutrino background after the energy and $\cos \theta_\odot$ cuts are obtained as
\begin{equation}
    \begin{aligned}
        N^{\rm geo} &= \sum_{i = 1}^{100~\rm{cells}} N_i^{\rm geo}, \\
        N^{\rm solar} &= \sum_{i = 1}^{100~\rm{cells}} N_i^{\rm solar}, \\
        N^{\rm rad} &= \sum_{i = 1}^{100~\rm{cells}} N_i^{\rm rad}, \\
    \end{aligned}
\end{equation}
and the total background uncertainty is determined by
\begin{equation}
    \begin{aligned}
        \sigma^{\rm bkg} = \big\{ & N^{\rm solar} + N^{\rm rad} + (N^{\rm solar} \times \sigma^{\rm solar})^2 \\
        &+ (N^{\rm rad} \times \sigma^{\rm rad})^2 \\
        &+ ((N^{\rm solar} + N^{\rm rad}) \times \sigma^{\rm cut})^2 \big\}^{1/2},
    \end{aligned}
    \label{eq;solarUncertainty}
\end{equation}
where $\sigma^{\rm solar}$, $\sigma^{\rm rad}$, and $\sigma^{\rm cut}$ are the relative systematic uncertainties of the solar neutrino flux, the radioactive purity, and the cut selection efficiency, respectively. 

The solar neutrino flux uncertainty, $\sigma^{\rm solar}$,
is assumed to be 1\% based on the expectation of a series of upcoming experiments or measurements, such as JUNO~\cite{JUNOSolar_2023}, Hyper K~\cite{HyperK_2024}, THEIA~\cite{THEIA_2020}, and JNE~\cite{JNE}.

The relative uncertainty of the selection efficiency, $\sigma^{\rm cut}$, is assumed to be 1\%. 
The efficiency of the fiducial volume cut and energy cut can be well studied by calibrations, thanks to the experiences gained by, for example, the Borexino~\cite{BorexinoCali_2012} and Daya Bay~\cite{DayaBayCali_2019, Zhangym} experiments.
The direction cut can be studied using a dedicated calibration source, a $\beta$-emitting source encapsulated in a metallic capsule with a collimated opening in one direction for the angular cut.

For the internal radioactive background, a 10\% uncertainty is used for $\sigma^{\rm rad}$. 
We expect two approaches can be used to measure it. 
First, assuming long-term decay equilibrium, it can be determined by measuring the event rate of Bi--Po cascade decays in the $\rm ^{238}U$ and $\rm ^{232}Th$ decay chains; with sufficient statistics, a 10\% uncertainty is achievable. 
The second approach is to fit the energy spectrum before the direction cut. The Borexino experiment has measured the solar Be7, pep, and CNO neutrino fluxes, as well as internal backgrounds~\cite{BorexinoPhase2_2019}.
The 10\% uncertainty is on the optimistic side.

Then, the sensitivity for geoneutrino detection is determined by
\begin{equation}
    {\rm Sensitivity} = \frac{N^{\rm geo}}{\sigma^{\rm bkg}},
\end{equation}
which estimates how likely the random fluctuations in the total background can reach the number of geoneutrino signals.

\begin{figure}
    \centering
    \includegraphics[width=\columnwidth]{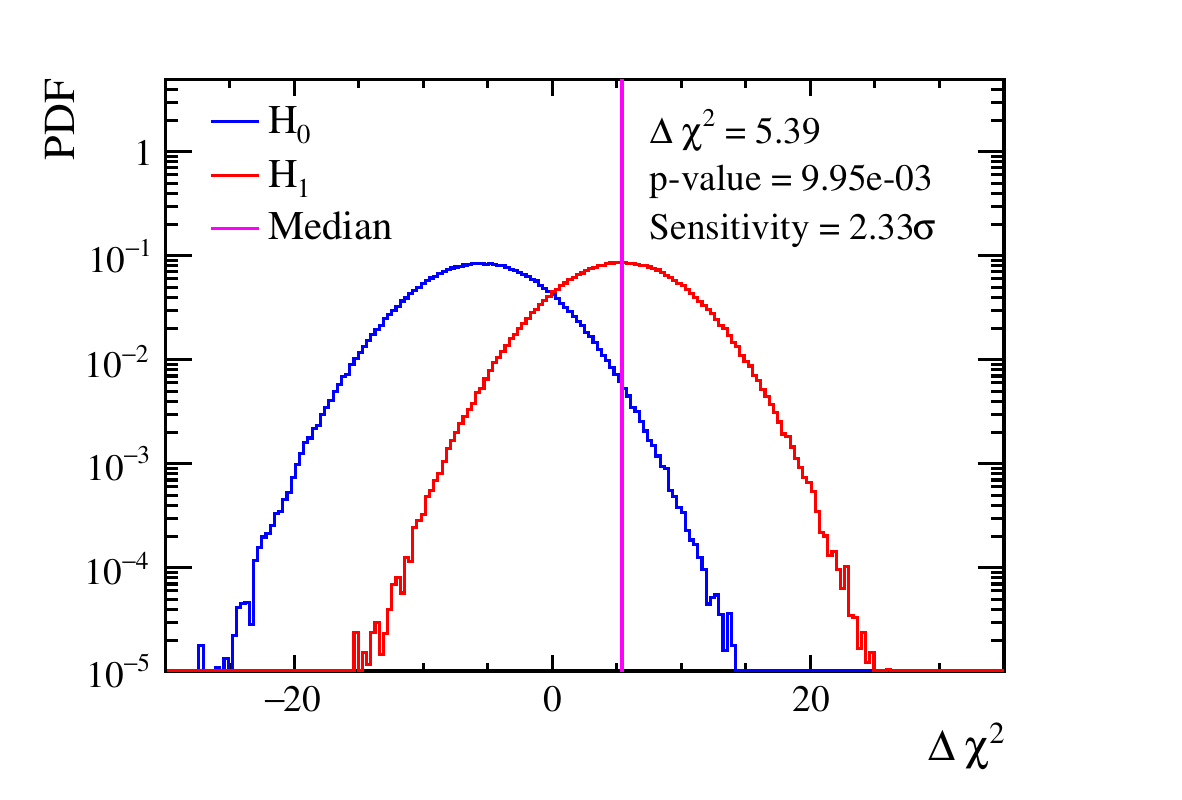}
    \caption{The test statistic $\Delta \chi^2$ distribution for two sample series under the two hypotheses $H_0$ (without geoneutrino) and $H_1$ (with geoneutrinos) at an exposure of 3 kiloton-years.}
    \label{fig:DeltaChi2}
\end{figure}

\begin{figure}[htp]
    \centering
    \begin{subfigure}{\columnwidth}
        \centering
        \includegraphics[width=\columnwidth]{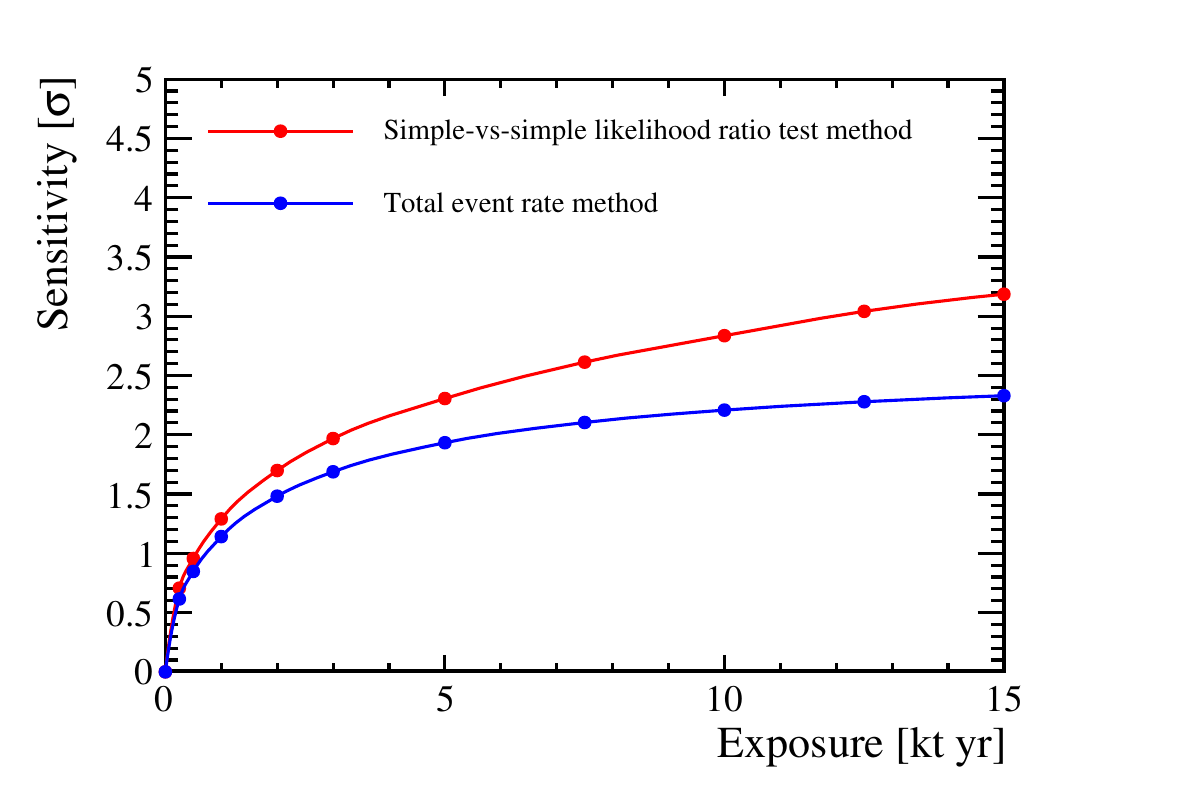}
        \caption{$\rm RoI_{Geo}$}
        \label{fig:Sensitivity_Geo}
    \end{subfigure}
    \begin{subfigure}{\columnwidth}
        \centering
        \includegraphics[width=\columnwidth]{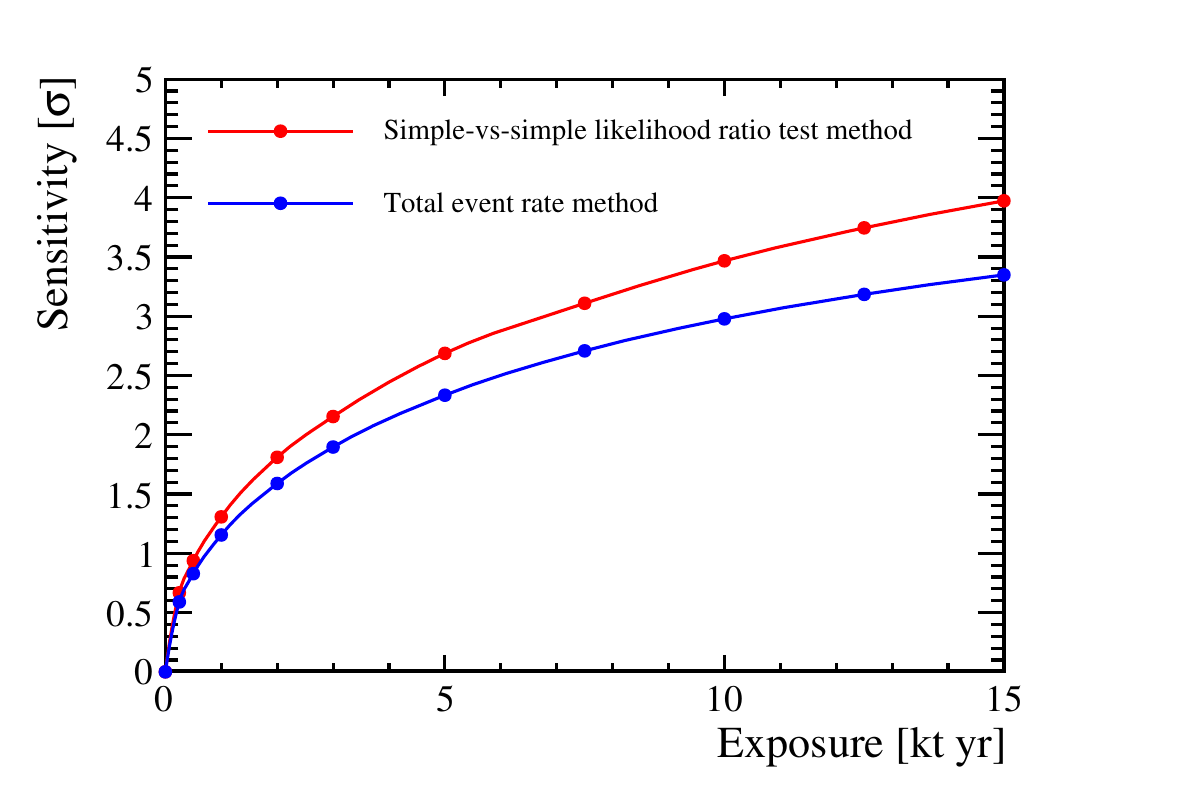}
        \caption{$\rm RoI_K$}
        \label{fig:Sensitivity_K}
    \end{subfigure}
    \begin{subfigure}{\columnwidth}
        \centering
        \includegraphics[width=\columnwidth]{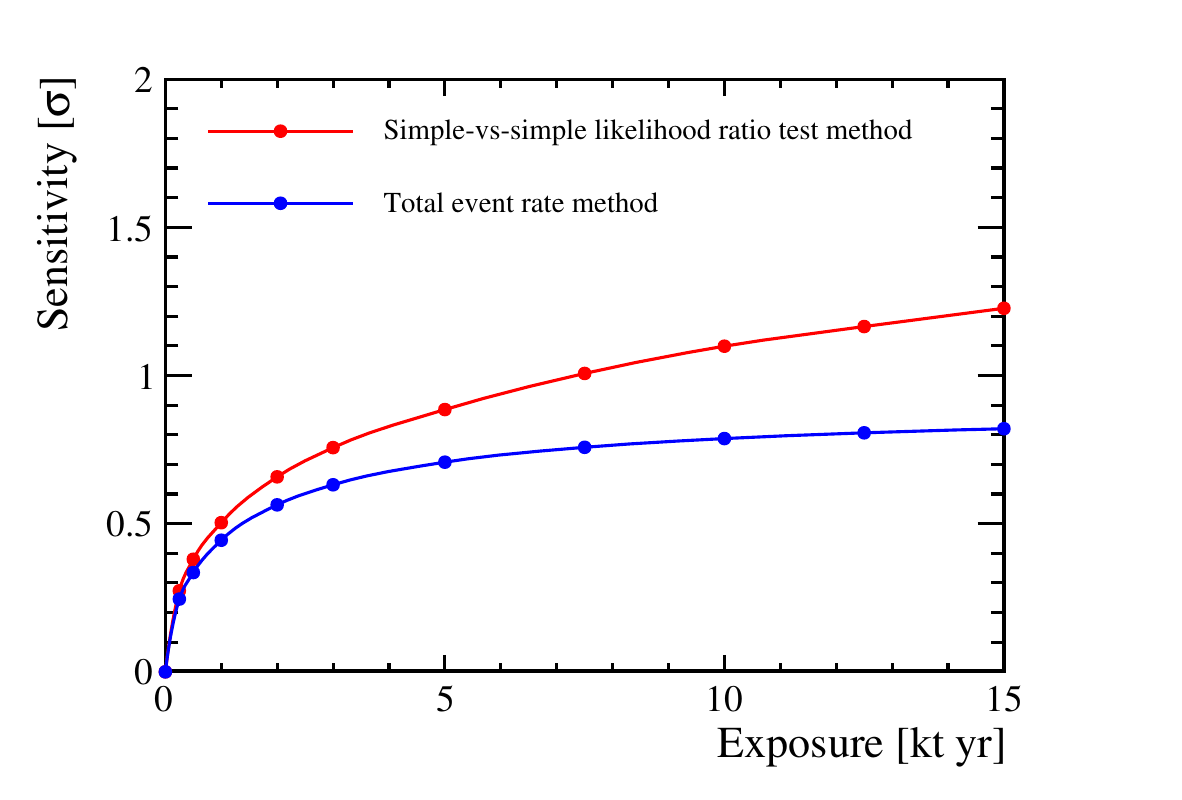}
        \caption{$\rm RoI_{U/Th}$}
        \label{fig:Sensitivity_ThU}
    \end{subfigure}
    \caption{Sensitivity to detect the geoneutrino signal in the three energy regions. The red solid line represents the sensitivity curve obtained using the simple-vs-simple likelihood ratio test method. In contrast, the blue solid line represents that obtained via the total event rate method.}
    \label{fig:Sensitivity_All}
\end{figure}

The second method accounts for the angular difference between the signals and backgrounds.
A simple-vs-simple likelihood ratio test is applied.
A statistic $\chi^2$ is defined based on the Poisson-form likelihood ratio with systematic uncertainties,
\begin{equation}
    \begin{aligned}
        \chi^2(\vec{n}) = {\rm min} \Bigg[ & -2 \sum_{i = 1}^{100~\rm{cells}} \left(n_i - \lambda_i + n_{i} \log \frac{\lambda_i}{n_i}\right) \\
        & + \left(\frac{\eta^{\rm solar}}{\sigma^{\rm solar}}\right)^2 + \left(\frac{\eta^{\rm rad}}{\sigma^{\rm rad}}\right)^2 + \left(\frac{\eta^{\rm cut}}{\sigma^{\rm cut}}\right)^2 \Bigg],
    \end{aligned}
    \label{equ:Chi2Define}
\end{equation}
where $n_i$ represents the number of candidates in the solid angle cell $i$ in a pseudo-experiment, $\lambda_i$ represents its prediction, and the three quadratic terms are the regularization terms introduced by the systematic uncertainties. 
Three pull parameters $\eta^{\rm solar}$, $\eta^{\rm rad}$, and $\eta^{\rm cut}$ are introduced to describe the deviations of the solar neutrino flux, the radioactive purity, and the cut efficiency from their nominal predictions, with uncertainties of $\sigma^{\rm solar}$, $\sigma^{\rm rad}$, and $\sigma^{\rm cut}$, respectively, as given in Eq.~\ref{eq;solarUncertainty}.
The prediction depends on the hypothesis $H_0$ (with no geoneutrino),
\begin{equation}
    \begin{aligned}
        \lambda_i = &\; N_i^{\rm solar} \times (1 + \eta^{\rm solar} + \eta^{\rm cut}) \\
        &+N_i^{\rm rad} \times (1 + \eta^{\rm rad} + \eta^{\rm cut}),
    \end{aligned}
\end{equation}
while for the alternative hypothesis $H_1$ with the predicted geoneutrino flux obtained in Sec.~\ref{sec:Signal},
\begin{equation}
    \begin{aligned}
        \lambda_i = &\; N_i^{\rm geo} + N_i^{\rm solar} \times (1 + \eta^{\rm solar} + \eta^{\rm cut}) \\
        &+ N_i^{\rm rad} \times (1 + \eta^{\rm rad} + \eta^{\rm cut}).
    \end{aligned}
\end{equation}
The simple-vs-simple test statistic is defined as
\begin{equation}
    \Delta \chi^2(\vec{n}) = \chi^2(\vec{n}; H_0) - \chi^2(\vec{n}; H_1).
    \label{eq:StatisticDefine}
\end{equation}
The sensitivity can then be obtained by calculating the $p$-value via a frequentist approach. Five million Monte Carlo datasets of both hypotheses are generated. The central value of the number of events in each cell is determined by sampling the systematic uncertainty, and the observed number of events is sampled according to Poisson statistics.
The distribution of $\Delta \chi^2$ at an exposure of 3 kiloton-years in $\rm RoI_{Geo}$ is shown in Fig.~\ref{fig:DeltaChi2}. 
The $p$-value, defined as the probability that the test statistic $\Delta \chi^2$ under the null hypothesis $H_0$ exceeds the median value of the $\Delta \chi^2$ distribution predicted by the hypothesis $H_1$, is calculated and subsequently converted into a sensitivity using the chi-squared distribution function. 
For the example of 3 kiloton-years exposure in Fig.~\ref{fig:DeltaChi2}, under the assumption that the null hypothesis $H_0$ is true, the $p$-value corresponding to the median value of the $\Delta \chi^2$ distribution predicted by the hypothesis $H_1$ is $9.95 \times 10^{-3}$, corresponding to a sensitivity of $2.33~ \sigma$.
A scan of sensitivity as a function of exposure is performed and presented in Fig.~\ref{fig:Sensitivity_All}.

The same analysis procedure is implemented for $\rm RoI_{K}$ and $\rm RoI_{U/Th}$. The sensitivity curves for the three energy regions are shown together in Fig.~\ref{fig:Sensitivity_All}.

For comparison, the previous result~\cite{Wang_2020} is also included. The significant improvement from the previous research to this study stems from the problem addressed in the detector response simulation in Sec.~\ref{sec:simulation}, the optimization of the $\cos \theta_\odot$ cut in each solid-angle cell, and the application of the simple-vs-simple likelihood ratio test method.
Specifically, to reach a $3~\sigma$ discovery sensitivity, the required exposure for the energy regions of
$\rm RoI_{Geo}$ and $\rm RoI_{K}$ is 11.9 and 6.9~kiloton-years, respectively.

\subsection{The Earth's large-scale structures imaging}

With directional reconstruction, the non-uniformity in the angular distribution of the geoneutrino signal can be observed, thereby enabling imaging of the large-scale structures of the Earth's interior. The process is demonstrated below.

\begin{figure}[htp]
    \centering
    \begin{subfigure}{\columnwidth}
        \includegraphics[width=\columnwidth]{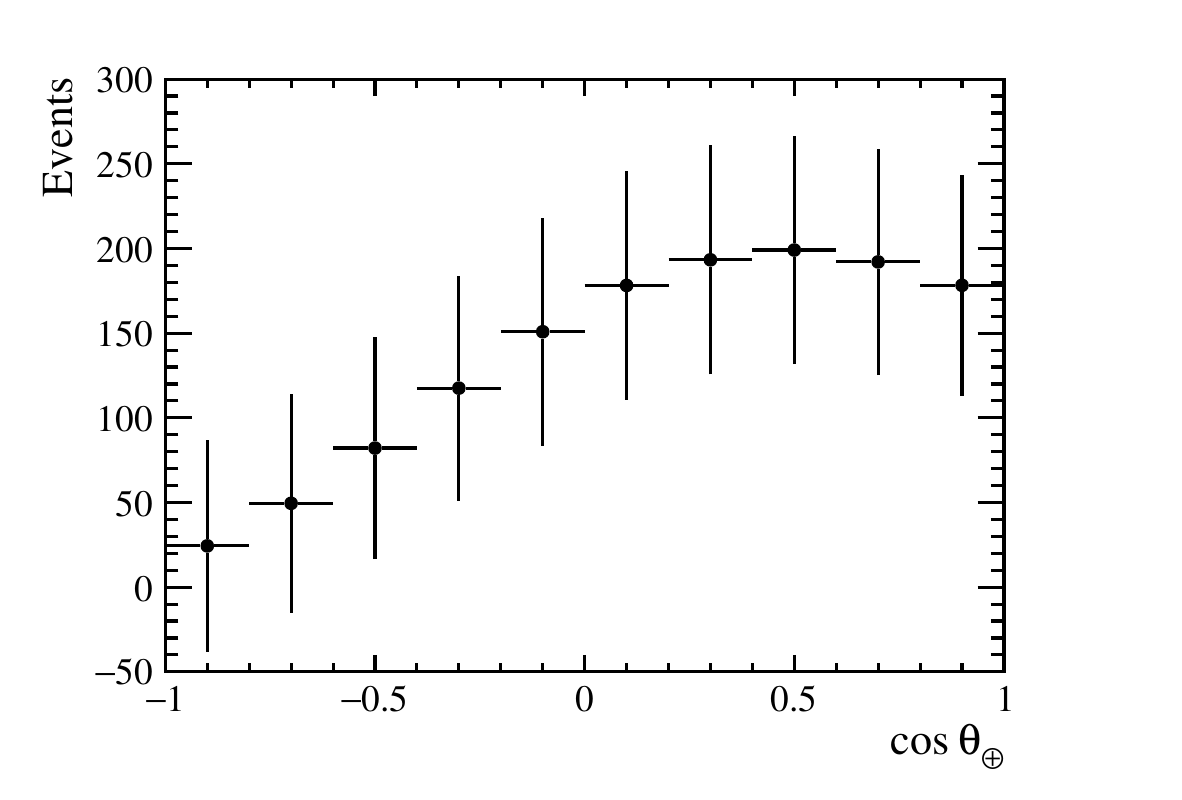}
        \caption{}
        \label{fig:Geo1DDirection_CosTheta}
    \end{subfigure}
    \begin{subfigure}{\columnwidth}
        \includegraphics[width=\columnwidth]{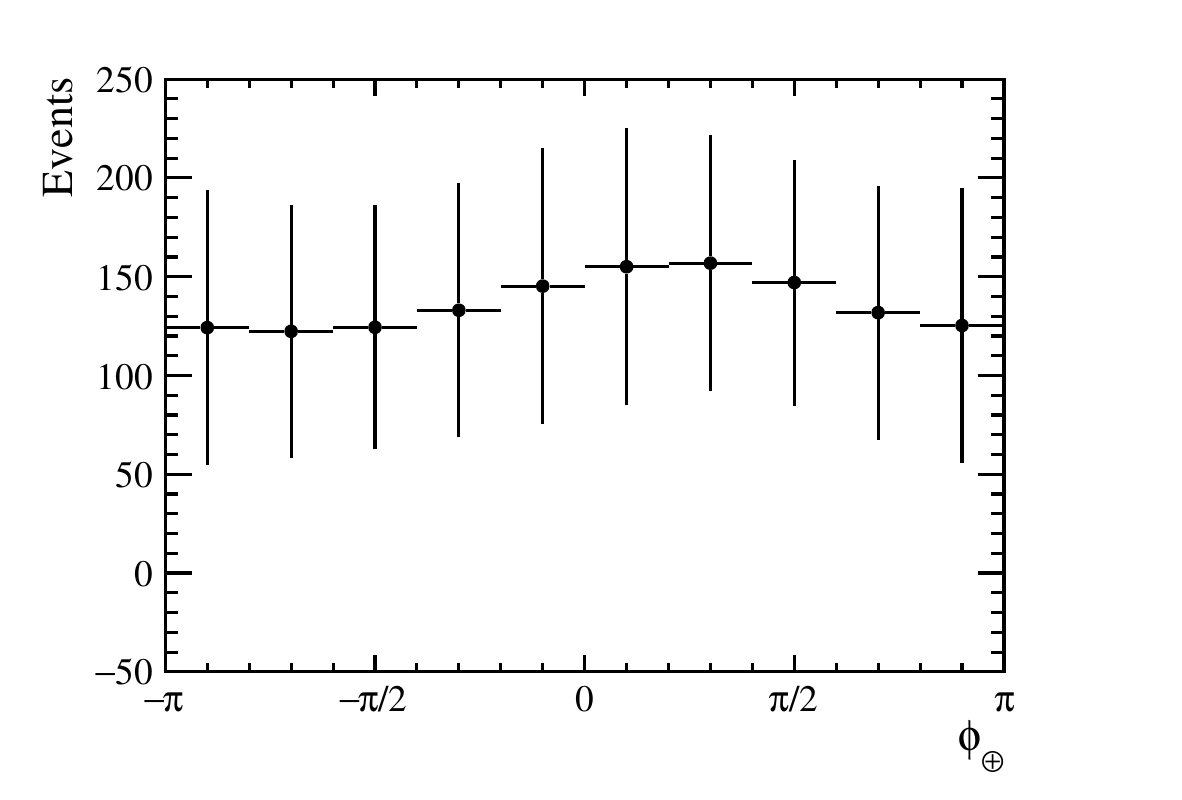}
        \caption{}
        \label{fig:Geo1DDirection_Phi}
    \end{subfigure}
    \caption{Projections of the geoneutrino signal angular distribution onto the $\cos \theta_\oplus$ (panel a) and $\phi_\oplus$ (panel b) directions at an exposure of 30~kiloton-years, with both the central values and error bars.}
    \label{fig:Projection}
\end{figure}

An Asimov dataset is generated. 
The geoneutrino signal is obtained by subtracting the solar neutrino and radioactive backgrounds from the candidates, as
\begin{equation}
    N_i^{\rm geo} = N_i^{\rm can} - N_i^{\rm solar} - N_i^{\rm rad},
\end{equation}
and its uncertainty is
\begin{equation}
    \begin{aligned}
        \sigma_i^{\rm geo} = \big\{ & (\sigma_i^{\rm can})^2 + (N_i^{\rm solar} \times \sigma^{\rm solar})^2 \\
        &+ (N_i^{\rm rad} \times \sigma^{\rm rad})^2 \\
        &+ ((N_i^{\rm solar} + N_i^{\rm rad}) \times \sigma^{\rm cut})^2 \big\}^{1/2}.
    \end{aligned}
\end{equation}
The raw geoneutrino signal's angular distribution, $N_i^{\rm geo, raw}$, is then obtained by correcting the cut selection efficiency, $\epsilon_i^{\rm geo}$, in each solid angular cell $i$ as
\begin{equation}
    N_i^{\rm geo, raw} = \frac{N_i^{\rm geo}}{\epsilon_i^{\rm geo}}.
\end{equation}
The projections of the raw geoneutrino signal's angular distribution onto the $\cos \theta_\oplus$ and $\phi_\oplus$ directions are produced to first reveal the non-uniform structure of the Earth, as shown in Fig.~\ref{fig:Projection}. 
An exposure of 30~kiloton-years is adopted in this demonstration. 
The concentration of the geoneutrino signal at $\cos \theta_\oplus > 0$ in Fig.~\ref{fig:Geo1DDirection_CosTheta} reflects the geoneutrinos' origin from the Earth,
while the bump in the azimuthal $\phi_\oplus$ direction in Fig.~\ref{fig:Geo1DDirection_Phi} is from the Qinghai-Tibet Plateau.

\begin{figure}[htp]
    \centering
    \includegraphics[width=\columnwidth]{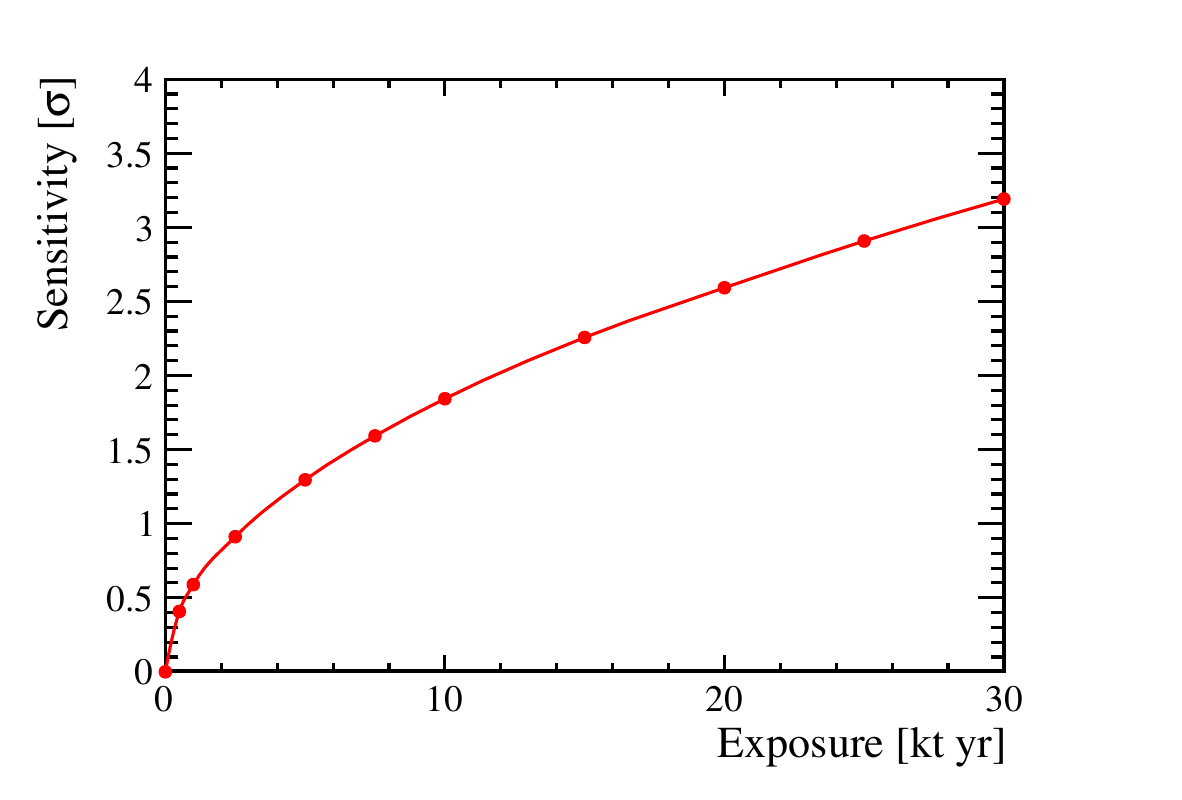}
    \caption{Sensitivity to detect the geoneutrinos' directional non-uniformity.}
    \label{fig:StructureSensitivity}
\end{figure}

The directional non-uniformity detection sensitivity is calculated via the same simple-vs-simple likelihood ratio test method as in Sec.~\ref{sec:GeoSens}. A uniform geoneutrino angular distribution with the same total flux is generated as the hypothesis $H_0$. The angular distribution obtained in Sec.~\ref{sec:Signal} serves as the hypothesis $H_1$. The likelihood and test statistic are defined the same as Eq.~\ref{equ:Chi2Define} and Eq.~\ref{eq:StatisticDefine}.
The distributions for both hypotheses are obtained through 5 million pseudo-experiments. The variation of directional non-uniformity detection sensitivity with exposure is shown in Fig.~\ref{fig:StructureSensitivity}. Specifically, to reach a $3~\sigma$ sensitivity, the required exposure is 27~kiloton-years.

\section{Discussion and summary}
\label{sec:Summary}
The potential to detect potassium-40 geoneutrinos and to image the Earth's large-scale structures is analyzed in this work. 
Using a combined Earth model that incorporates the CRUST1.0 crust model and a medium-Q mantle model, the event rate of geoneutrino-electron elastic scattering is calculated.
The critical solar neutrino 
and internal radioactive backgrounds are also predicted.

A Cherenkov liquid scintillator detector is implemented to simulate the recoil electrons of neutrinos. 
A problem in the previous study has been resolved, and the detector's angular response has been derived.
Geoneutrinos are mainly from the sediments and upper crust, while the intrinsic background, solar neutrinos, is aligned with the Sun's position. 
The full $4 \pi$ space in the terrestrial coordinate system is divided into 100 solid-angle cells, and an optimized solar angle cut is achieved independently for each cell by maximizing the local signal-to-background ratio.

The geoneutrino detection sensitivity is improved.
The $3~\sigma$ sensitivity to discover geoneutrinos in the energy region dominated by potassium corresponds to an exposure of 6.9~kiloton-years.
With the direction information, an image of the Earth’s large-scale structures can also be made, and the required exposure is 27~kiloton-years to reject a uniform geoneutrino distribution by $3~\sigma$.

The detection efficiency (i.e.~photocathode coverage and quantum efficiency of PMT) of optical photons of the detector is a critical factor to achieve the required angular resolution. 
The radioactive background level and the determination of its uncertainty are another key issue to reach the desired sensitivity.
A larger detector with a larger fiducial volume, such as SNO+, can certainly accelerate the experimental study.

\section*{Acknowledgments}
This work is supported in part by 
the Ministry of Science and Technology of China (No. 2022YFA1604704),
the National Natural Science Foundation of China (No. 12141503),
the Key Laboratory of Particle \& Radiation Imaging (Tsinghua University), and the CAS Center for Excellence in Particle
Physics (CCEPP).

\bibliographystyle{cas-model1-num-names}
\bibliography{GeoNeutrino}

\end{document}